\documentclass{JHEP}

\usepackage{amsmath}
\usepackage{epsfig}
\usepackage{amsfonts}
\usepackage{amssymb}
\usepackage{latexsym}

\newcommand{\Mn}{m_n}

\def\sqr#1#2{{\vbox{\hrule height.#2pt\hbox{\vrule width.#2pt
height#1pt \kern#1pt \vrule width.#2pt}\hrule height.#2pt}}}
\def\QQ{\mathchoice\sqr54\sqr54\sqr43\sqr33}

\def \lsim{\mathrel{\vcenter{\hbox{$<$}\nointerlineskip\hbox{$\sim$}}}}
\def \gsim{\mathrel{\vcenter{\hbox{$>$}\nointerlineskip\hbox{$\sim$}}}}

\newcommand{\mud}{m}%{{\dot{\mu}}}
\newcommand{\nud}{n}%{{\dot{\nu}}}
\newcommand{\alphad}{i}%{{\dot{\alpha}}}
\newcommand{\betad}{j}%{{\dot{\beta}}}

\def\Tcontr#1#2#3#4{\raise#2\rlap{\hskip#1
\hbox{\vrule width 0.4pt height#3}%
\raise#3\hbox{\vrule width #4 height 0.4pt}%
\hbox{\vrule width 0.4pt height#3}}}

\newcommand{\GeV}{\,{\rm GeV}}
\newcommand{\TeV}{\,{\rm TeV}}

\newcommand{\eq}[1]{~(\ref{eq:#1})}
\def\Tr{\mathop{\rm Tr}}
\def\circa#1{\,\raise.3ex\hbox{$#1$\kern-.75em\lower1ex\hbox{$\sim$}}\,}

\title{Graviton loops and brane observables}

\author{Roberto Contino$^1$, Luigi Pilo$^1$, Riccardo Rattazzi$^{1,2}$, 
Alessandro Strumia$^2$\thanks{also at Dipartimento di Fisica
dell'Universit\`a di Pisa and INFN.}
 \\ $^1$Scuola Normale Superiore, Piazza dei Cavalieri 7, I-56126 Pisa, Italy $\&$ INFN
 \\[0.3cm] $^2$Theory division, CERN, CH-1211 Geneva 23, Switzerland}

\abstract{ We discuss how to consistently perform effective Lagrangian 
computations in quantum gravity with branes in compact extra dimensions.
A reparametrization invariant and infrared finite result is obtained in a non trivial way.
It 
is crucial to properly account for brane fluctuations and to correctly identify
physical observables. Our results correct some confusing claims in
the literature.
We discuss the implications of graviton loops on electroweak
precision observables and on the muon $g-2$ in models with large extra dimensions.
We model the leading effects, not controlled by effective
field theory, by introducing a hard momentum cut-off.}

\preprint{CERN--TH/2000--383 \\ IFUP--TH/2000--24 \\ SNS-PH/01--04}
\keywords{Extra Dimensions, Gravity, Branes, LEP physics}

\begin{document}

\newpage

\section{Introduction}
\def\baselinestretch{1}

No known experimental constraint firmly excludes the possibility that
Kaluza Klein (KK) excitations of the graviton propagating in $\delta\geq 2$ 
large extra dimensions
will affect future particle physics experiments~\cite{large,large2}\footnote{
The case $\delta=2$ is excluded by bounds on emission of KK modes with a
small mass $\lsim 100$ MeV in supernovae~\cite{large,nova}, if the extra dimensions are flat.
In principle, one could save collider signals (due to heavy KK modes)
by assuming that the compact dimensions are  curved on length scales $\gsim (100 {\rm MeV})^{-1}$,
so that the light KK are lifted.}.
After removing all non propagating 
degrees of freedom by a suitable choice of coordinates,
many authors computed the signals of KK graviton emission at tree level~\cite{GRW,KKgr,KKgr2}.
Some authors also considered 1-loop effects~\cite{KKgr2,EWPOgr1,EWPOgr2,Bij,grasser}:
since they affect observables measured with higher precision, 
they can compete with tree level effects.
The result was not the expected one.
Consider for example the graviton correction to the Higgs mass.
At first sight one would estimate it as
\begin{equation}\label{i1}
\delta m_h^2 \sim \sum_n \int_k \frac{(m_h k/M_4)^2}{(k^2 + \Mn^2) (k^2+m_h^2)}
\sim \sum_n \frac {m_h^2 \Lambda^{2}}{(4\pi)^2 M_4^2} \sim  m_h^2 \bigg(\frac {\Lambda}{ M_D}\bigg)^{\!\! 2+\delta}
\end{equation}
where $M_D$, $M_4$ are the gravitational scales of the $D$-dimensional and $4$-dimensional theory
respectively, and $\Lambda \sim M_D $ parameterizes the unknown quantum gravity ultraviolet cutoff.
At a closer look \cite{KKgr2,EWPOgr1,EWPOgr2,Bij}
the effect seems to be much larger.
To understand that, consider the propagator for the physical $J=2$ 
$n$th KK graviton with mass $\Mn$ and 4-momentum $k_\mu$~\cite{GRW,KKgr2}
\begin{equation}\label{eq:GG}
\Tcontr{0.5em}{2.5ex}{1ex}{1.7em} G_{\mu\nu}^{(n)}
G_{\rho\sigma}^{(m)}= \frac{i\delta_{n,-m}}{(k^2-\Mn^2)} \, \frac{1}{2}
 (t_{\mu\rho}t_{\nu\sigma}+t_{\mu\sigma}t_{\nu\rho}-\frac{2}{3} t_{\mu\nu}t_{\rho\sigma})
\end{equation}
where $t_{\mu\nu}\equiv \eta_{\mu\nu}- k_\mu k_\nu/\Mn^2$.
If the terms enhanced by powers of $k/\Mn$  were to fully    contribute 
to quantum corrections,
the $k$ factors would give a highly ultraviolet (UV) divergent loop effect.
More importantly, when $\delta<4$ the $1/\Mn$ factors would also give a strong 
infrared (IR) enhancement of the sum over KK modes.
At the end the correction would be a factor $(M_D R)^{4-\delta}$ larger 
than the naive one in eq.~(\ref{i1}), where $R$ is the size of the extra 
dimensions.
This kind of behavior, indeed observed in~\cite{KKgr2,EWPOgr1,EWPOgr2,Bij},
would exclude the possibility that $\delta < 4$ large extra dimensions
(i.e.\ $R\gg 1/M_D$) solve the hierarchy problem. 

The above argument on the fate of the $k/\Mn$ terms must however be wrong,
and for a very simple reason. Indeed one could choose to fix the 
$D$-dimensional reparametrization invariance by the
de Donder gauge choice, in which the graviton propagator contains no 
$k/\Mn$ terms. This is in complete analogy with the case of a massive vector boson,
where the propagator contains  $k/\Mn$ terms in the unitary gauge,
while no such term is present in the Feynman gauge.
Therefore $k/\Mn$ terms cannot affect gauge-invariant physical observables.
This suggests that there must be something
missing or incorrect in the  computations so far performed.

The purpose of the present paper is to devise all the elements that are needed 
for a fully consistent computation. The guideline is to respect the full
$D$-dimensional general coordinate covariance. First of all it is crucial
to fix the gauge by the Faddeev-Popov procedure and to choose
a covariant regulator. If the regulation of the loop integral is
not performed with the due care, spurious UV and IR divergences can
appear. Secondly, one has to remember that the position of the brane
depends on the system of coordinates, and therefore brane fluctuations ({\em branons}) 
must be taken into account in order to respect general covariance. Finally
one must carefully identify  which are the physical observables  in
the presence of gravity: misidentification of the true observables
can yield spurious gauge dependence and IR divergences.
 
One of our results is that all the puzzling effects
found in the existing literature cancel out in a fully
consistent calculation when one computes physical
observables. 
For example the Higgs mass term and the oblique $S,T,U$ parameters~\cite{STU} 
are not physical observables (except in particular cases).
So they
receive gauge dependent quantum gravity corrections, which in some cases are even
 enhanced by powers of $RM_D$. These infrared pathologies, which would
invalidate perturbation theory (for instance $R M_D\sim 10^{15}$ if $R\sim {\rm mm}$ and $M_D\sim \TeV$),
are absent in the
corrections that affect the corresponding physical observables,
the pole higgs mass and the $\epsilon_1, \epsilon_2, \epsilon_3$ parameters~\cite{epsilon}.

In our study we treat quantum gravity and the brane by the method of effective 
field theory (EFT)~\cite{Dono,sundrum}. 
 We do so in the absence of a  realistic fundamental description
 of the SM on a brane\footnote{Interesting attempts based on $D$-brane intersections~\cite{ibanez}
 give `semi-realistic' models with extra charged
matter with  respect to the SM. The stability of these configurations is an open question.}.
 The effective Lagrangian  summarizes all our 
low energy  knowledge of gravitational interactions with SM particles.
By our method we could perform a fully consistent computation of the
1-loop quantum gravity corrections to electroweak precision
observables.
However the dominant effects are strictly speaking
uncalculable, as they are saturated in the UV where we loose
control of the theory.
We can only parameterize these effects in terms of a UV cutoff $\Lambda$\footnote{A string model could provide a
physical realization of this cut off. However at the level of the present model
building technology there are many free parameters specifying the moduli
and the brane configuration~\cite{ibanez,Aldazabal:2000sa,Antoniadis:2001bh}. 
Therefore, even if we were able to reproduce the SM,
the predictive power on quantum corrections (for example on the muon $g-2$) 
would probably be limited. Of course it
would still be important to have one such model. Indeed it would also be interesting
to have a field theoretic brane model in the spirit  of~\cite{Rubakov,Dvali}.}.
The calculable piece is the one saturated in the infrared, but this
is only of order $(M_Z/M_D)^{2+\delta}$. 
Therefore, introducing a UV cut-off $\Lambda$, we will only compute a particular
combination of observables, which is affected by just a few simple
Feynman diagrams. For the full set of observables we will
limit ourselves to a qualitative discussion.

While the discussion of the phenomenology is somewhat limited by the powerlike
UV divergences, we stress that the main goal of the present paper is 
conceptual. In this respect the most important (and new)  result is that
brane motions have to be properly taken into account. In order
to understand this issue better we have also considered the case of
a brane living at an orbifold fixed point, for which the branons are
projected out. 
In this case gauge independence of observables is met through tadpole diagrams specific of orbifold
compactifications, rather than by branon loops. 
The technology developed in this paper may prove useful in future work.
One possible application is the brane to brane mediation of supersymmetry
breaking through bulk gravity at 1-loop. This effect
is computable and represents the leading correction to anomaly mediated soft terms:
depending on its sign it may cure the tachyon
problem of anomaly mediation. 

The paper is structured as follows.
In section 2 we discuss our Lagrangian and the effective field theory
philosophy. We also introduce various gauge fixing conditions for
the gravitational field and explain the r\^ole of the branons.
In section 3 we calculate the corrections to the masses
of scalars and vectors, on and off the brane. We explain what gauge independence
means in quantum gravity, and show that physical quantities are gauge independent.
We also give the example of a brane at an orbifold fixed point, for
which branons are not needed.
In section 4 we derive experimental bounds
on low energy quantum gravity from precision measurements
and from the anomalous magnetic moment of the $\mu$. In
section 5 we summarize.  Finally in the appendices we describe how to derive graviton-matter
vertices and collect our results for the corrections to brane observables.

\section{Effective Lagrangians for gravitons and branes}

\subsection{Pure gravity} \label{Pure gravity}

We study gravity in $\mathbb{R}^d \times {\cal M}$ where the extra dimension ${\cal M}$ is a
compact manifold of dimension $\delta$.
Not knowing which manifold is of physical interest (if any), we consider the simplest one: a 
$\delta$-torus $T^\delta$ with a
single radius $R$ and
volume $V=(2\pi R)^\delta$. 
We perturbatively expand the classical Einstein-Hilbert action around the flat metric 
$g_{MN}=\eta_{MN}+\kappa \, h_{MN}$, $\, \eta_{MN}=(+1,-1,-1,-1,\dots)$
 in terms of the graviton field $h_{MN}$:
\begin{equation} \label{eq:linear}
\begin{split}
{\cal S} &= \frac{\bar M^{D-2}_D}{2} \int d^DX \sqrt{g} \; R \\
 &= \frac{1}{2} \int d^DX \; \Big(
 - h^{MN} \Box h_{MN} + h \Box h  -2 h^{MN} \partial_{M} \partial_{N} h
 +2 h^{MR} \partial_{R} \partial_{S} h^S_{\;M} \Big) +{\cal O}(\kappa).
\end{split}
\end{equation}
where $D=d+\delta$, $h\equiv h_M^M$ and we used $\eta_{MN}$ to raise and lower indices.
We use upper (lower) case latin letters for $D$-dimensional
(extra-dimensional) indices and Greek letters for $d$-dimensional indices; in particular we decompose
the $D$-dimensional coordinates as $X^M=(x^\mu,\,y^i)$.
We do not fix $d=4$ since we will use dimensional regularization.
Following  the notation of ref.~\cite{GRW} we have defined
\begin{equation} \label{md}
\kappa^2 \equiv 4 \bar M_D^{2-D}\, , \qquad \bar M_d^{d-2} = V \bar M_D^{D-2} = R^\delta M_D^{D-2}
\end{equation}
$\bar M_d$ is the effective reduced Planck mass as measured by a 
$d$-dimensional observer, $\bar M_D$ is the corresponding parameter in 
$D$ dimension and  $M_D$ is defined by~(\ref{md}).
With this convention the  equations of motion read:
\begin{equation}
R_{MN} - \frac{1}{2} g_{MN} R = - \frac{1}{\bar M_D^{D-2}} T_{MN} 
\end{equation}
Before inverting the quadratic term in eq.~(\ref{eq:linear}) to obtain the propagators,
one must fix the reparame\-tri\-zation invariance;
we follow the Faddeev-Popov procedure and introduce a set of $\xi$-gauges by
adding to the Lagrangian the gauge-fixing term ${\cal L}_{\rm GF}=-F^2/\xi$, where
\begin{equation} \label{eq:GF}
F_N = \partial^\mu (h_{\mu N}-\frac{1}{2}\eta_{\mu N} h) +
 \xi  \partial^i (h_{iN}-\frac{1}{2\xi}\eta_{iN} h).
\end{equation}
This particular choice breaks the $D$-dimensional Lorentz symmetry of the flat 
background metric
for generic values of $\xi$ and
interpolates between the usual de Donder and unitary gauge, obtained respectively
in the limit $\xi\to 1,\infty$. The functional
integral gets multiplied by the  Faddeev-Popov determinant, 
exponentiated in the usual way by introducing `ghost' fields $\eta_M$, $\bar \eta_M$:
\begin{equation}
{\cal L}_{\rm ghost}= \int d^DX \, d^DX' \; \bar \eta_N(X) 
 \frac{\delta F_N(X)}{\delta \lambda_M(X')}\Big|_{\lambda=0} \eta_M(X')
\end{equation}
where $\lambda$ is the gauge parameter for reparametrizations.

The kinetic term for the graviton field is a (messy) $3\times 3$ matrix which mixes 
tensor $h_{\mu\nu}$, vector $h_{\mu i}$ and scalar $h_{ij}$ modes. 
Since interactions are more easily written in terms of the $h_{\mu\nu}$, $h_{\mu i}$ and $h_{ij}$
components of the $D$-dimensional graviton field $h_{MN}$,
it is more convenient to write the propagator in this basis rather than in the gauge-dependent mass eigenstate basis.
For example matter fields confined on a straight
$d$-dimensional brane at leading order  couple only to the tensorial $h_{\mu\nu}$ mode.

\smallskip

By decomposing the graviton field $h_{MN}$ in its Fourier harmonics
\begin{equation} \label{eq:fourier}
h_{MN}(x,y) = \frac{1}{\sqrt{V}}  \sum_{n\in \mathbb{Z}^\delta} h^{(n)}_{MN}(x)\,  e^{in\cdot y/R}
\end{equation}
and integrating the Einstein-Hilbert Lagrangian over the extra-coordinates,
one obtains the $d$-dimensional Lagrangian for KK modes.

Notice that $\partial^i h_{iN}$ can be interpreted as Goldstone bosons eaten in
 a gravitational Higgs mechanism to form massive tensors and vectors. We are classifying
particles by the d-dimensional Poincar\`e group.  By this interpretation, eq.~(\ref{eq:GF})
is the analogue of 't Hooft's $\xi$ gauge in spontaneously broken gauge theories.
For $\xi\to\infty$ we get the unitary gauge \cite{GRW,KKgr2}
in which only the physical degrees of freedom propagate.
In this limit the $\xi$ gauge propagator for the modes 
$h_{MN}=(h_{\mu\nu}$, $h_{ij}$, $h_{\mu j})$ simplifies to
\begin{equation}\label{eq:PU}
\begin{split}
\Tcontr{0.5em}{2.5ex}{1ex}{2.4em} h_{MN}^{(n)} &h_{M'N'}^{(n')}  =
\frac{1}{2}\frac{i\delta_{n,-n'}}{(k^2-\Mn^2)}\cdot
\\ 
 & \begin{bmatrix} 
 t_{\mu\mu'}t_{\nu\nu'}+t_{\mu\nu'}t_{\mu'\nu}-
  \frac{2}{D-2} t_{\mu\nu}t_{\mu'\nu'} & \frac{2}{D-2} P_{i'j'} t_{\mu\nu} & 0 \\
 \frac{2}{D-2} P_{ij} t_{\mu'\nu'} & 
P_{ii'}P_{jj'}+P_{ij'}P_{ji'}-\frac{2}{D-2} P_{ij}P_{i'j'} & 0 \\
 0 & 0 & -P_{ii'} t_{\mu\mu'} 
\end{bmatrix}
\end{split}
\end{equation}
where $k$ is the $d$-dimensional momentum,
\begin{equation}
P_{ij} \equiv \delta_{ij}-\frac{n_i n_j}{n^2},\qquad  t_{\mu\nu} \equiv \eta_{\mu\nu}-\frac{k_\mu k_\nu}{\Mn^2}
\end{equation}
and $\Mn^2=n^2/R^2 $ is the mass squared for the $n$th KK excitation, having defined
$n^2\equiv -n_i n_j \eta^{ij}=n_i n_j \delta_{ij}$.
In appendix~A we derive this propagator by working in the unitary gauge with physical fields.
As usual `Goldstone' bosons and `ghosts' get infinitely massive when $\xi\to\infty$ but they
do not decouple:
loop corrections computed in the unitary gauge ($\xi=\infty$) by propagating
only the physical fields are different 
from the limit $\xi\to\infty$ of loop effects computed in a $\xi$ gauge \cite{pasarin}. Of course
the mismatch disappears in physical quantities.

\medskip

For $\xi=1$ we get instead the de Donder gauge, where the propagator has the covariant form:
\begin{equation}
\Tcontr{0.1em}{2.0ex}{1ex}{2.4em} h_{MN}h_{M'N'} = \frac{i}{2K^2}  (\eta_{MM'}\eta_{NN'}+\eta_{MN'}\eta_{NM'}-\frac{2}{D-2} \eta_{MN}\eta_{M'N'})
\end{equation}
where $K$ is the $D$-dimensional momentum.
In matrix notation, for the single KK mode:
\begin{equation}\label{eq:PDD}
\begin{split}
\Tcontr{0.5em}{2.5ex}{1ex}{2.4em} h_{MN}^{(n)} &h_{M'N'}^{(n')} =  \frac{1}{2}\frac{i\delta_{n,-n'}}{(k^2-\Mn^2)} \cdot
\\
 &\quad \begin{bmatrix} 
\eta_{\mu\mu'}\eta_{\nu\nu'}+\eta_{\mu\nu'}\eta_{\nu\mu'}-
  \frac{2}{D-2} \eta_{\mu\nu}\eta_{\mu'\nu'} &
 \frac{2}{D-2} \delta_{i'j'} \eta_{\mu\nu} & 0 \\
 \frac{2}{D-2} \delta_{ij} \eta_{\mu'\nu'} & 
\delta_{ii'}\delta_{jj'}+\delta_{ij'}\delta_{ji'}-
  \frac{2}{D-2} \delta_{ij}\delta_{i'j'} & 0 \\
 0 & 0 & -\delta_{ii'} \eta_{\mu\mu'} 
\end{bmatrix}
\end{split}
\end{equation}
We have thus shown that the propagator in the de Donder and unitary gauges
 has the same form up to longitudinal $k/\Mn$ terms.
For compactness, we do not write explicitly the propagator in a generic $\xi$-gauge.

\subsection{Gravity and branes} \label{brane}

We want to study quantum gravity corrections to the physical observables
of a field theory living on a $d$-dimensional brane  in a
$\delta$-dimensional compact space $\cal M$.
At the end we will identify the brane theory with the Standard Model.
The gravitational Lagrangian has been discussed in the previous subsection.
We now discuss the brane Lagrangian.

We  use an effective field theory (EFT) approach where the fundamental description
of the particles and of the brane is not specified \cite{sundrum}. 
In the regime of validity
of EFT, the particles are treated as point-like and
the brane is treated as infinitely thin in the extra dimensions.
This requires a little explanation. 
If $\rho$ is the brane true transverse size, our EFT is only valid at energy scales
$\ll 1/\rho$. The brane  also generally has a finite tension $\tau \equiv f^d$.
This gives rise to a gravitational field behaving like $f^d/M_D^{d+\delta-2} 
r^{\delta-2} \equiv
(r_G/r)^{\delta-2}$ at a distance $r$ far away from the brane in the extra space.
We focus on $\delta>2$ (for $\delta=2$ the background is locally flat
with a conical singularity at the brane position).
The gravitational radius $r_G$ controls the distance at which the geometry is curved. 
One can then think of different possibilities for the brane structure.
If $\rho>r_G$ the brane is similar to a big star where the geometry nowhere strongly 
deviates from the flat, and $1/\rho$ truly represents the UV cut-off of our EFT.
On the other hand for $\rho \ll r_G$ it is the gravitational radius that sets the
UV cut-off. Physics at energies $>1/r_G$ would probe the gravitational structure
of the brane, which is non-universal and model dependent. One example
is a black brane where at $r\sim r_G$ a black-hole horizon is present. Another
different example is given by the solution studied in~\cite{charmousis}, 
where there is no horizon and a naked singularity is avoided by a finite brane size.
In the latter case the coupling of bulk gravitons to the brane is dramatically
changed at energies $>1/r_G$. As we are only interested in universal features
we will assume that the UV cut off $\Lambda_{\rm UV}$ that limits the use of our EFT
is bounded by ${\rm min}(1/\rho,1/r_G)$. In the regime of validity of EFT
we can treat the background metric as approximately flat and treat the 
effects of the brane tension as  perturbations.
 Notice indeed that at energy $E$ the latter are controlled by  the small parameter
$E^{\delta-2} f^d/M_D^{d+\delta-2}\equiv (E r_G)^{\delta-2}<(\Lambda_{\rm UV} r_G)^{\delta-2}\ll 1$ .

Two possibilities are given: either the brane 
can freely move in the bulk or sit at a fixed point, if the compact
space ${\cal M}$ has any.
Let us consider the former case first. The immersion
of the brane in the $D$-dimensional space is parameterized by $D$ functions $X^M(z)$,
where $z^\mu$ are the $d$ local coordinates on the brane. The brane
action must be invariant under both $D$-dimensional 
coordinate changes (under which $X^M$ transform and $z^\mu$ are unchanged)
and under reparametrizations of the brane coordinates $z^\mu$.
An invariant brane action can be built using the induced metric 
\begin{equation}
g_{\mu\nu}^{\rm ind}(z) = \frac{\partial X^M(z)}{\partial z^\mu}
\frac{\partial X^N(z)}{\partial z^\nu} g_{MN}(X(z)).
\end{equation}
Since $g^{\rm ind}_{\mu\nu}$ is a scalar under $D$-dimensional reparametrizations, 
we only need to respect brane reparametrizations
by the use of $g^{\rm ind}_{\mu\nu}$ itself. The description of the brane position
by the $X^M(z)$ is of course redundant. We can eliminate this redundancy
by using the remaining gauge freedom represented by brane reparametrizations.
We stress that we cannot use $D$-dimensional diffeomorphisms for which the
gauge has been completely fixed in the previous section.
A convenient choice of brane coordinates is $x^\mu=z^\mu$, 
$y^i=\xi^i(x^\mu)$. This choice completely fixes brane reparametrizations without
the need of introducing additional  ghost fields (the ghost determinant is trivial)~\cite{sundrum}. 
We call the $\xi^i$ {\it branons}.

As we said the branons cannot be thrown away because we have already completely fixed
the $D$-dimensional reparametrization gauge invariance. However in the previous section one could 
have chosen a different class of coordinate gauges, one in which the brane always sits
at a given point in ${\cal M}$. This different choice would explicitly 
break translation invariance in the extra dimensions. What becomes of the branons
in these different gauges? They are still there but as longitudinal modes of a
combination of graviphotons: the branons can indeed be interpreted as the Goldstone 
bosons of  broken translation invariance in the extra dimensions~\cite{sundrum}.
We find it 
more convenient to gauge fix the graviton in the more standard way and keep the branons.
Notice that, consistently with their Goldstone character, in the limit in which gravity 
decouples ($M_D\to \infty$) the  branons survive. Their physical effects can  therefore
 be studied  independently of gravity~\cite{branoni}. Quantum fluctuations of the branons
are controlled by $1/\tau$ (the analogue of $1/f_\pi^2$ for pions) and
become non-perturbative at an energy $E>{\sqrt {4\pi}} f$ ($E>4\pi f_\pi$ for pions). 
Therefore 
the tension $\tau$ sets another sure upper bound on the regime of applicability of EFT.

In terms of the branons $\xi^i$ the induced metric is 
\begin{equation}\label{eq:heff}
g_{\mu\nu}^{\rm ind} = 
g_{\mu\nu} -g_{\mu i}g_{\nu j} g^{ij}+(D_\mu \xi_i)(D_\nu \xi_j ) g^{ij}\equiv 
\eta_{\mu\nu}+\tilde{h}_{\mu\nu}
\end{equation}
where $D_\mu \xi_i \equiv \partial_\mu \xi_i + g_{\mu i}$ and the metric $g_{MN}$ is
evaluated at the brane location $y^i=\xi^i$. For 1-loop computations we need 
$\tilde h_{\mu\nu}$ up to quadratic order in $\xi$
\begin{equation}
\tilde h_{\mu\nu}=\kappa h_{\mu\nu}+(\partial_\mu \xi_i) (\partial_\nu \xi^i)+
\kappa(\xi^i \partial_i h_{\mu\nu}+ h_{i\mu} \partial_{\nu} \xi^i+ h_{i\nu} \partial_{\mu} \xi^i)+\cdots
\end{equation}
where now $h$ is the graviton field evaluated at the brane rest position $y^i=0$.
The brane Lagrangian is given by
\begin{equation}\label{eq:Lbrane}
{\cal S}_{\rm brane} = \int d^4 x  \bigg[-\tau \sqrt{\det g_{\mu\nu}^{\rm ind}}  + {\cal L_{\rm SM}}
+\cdots\bigg]
\end{equation}
where ${\cal L_{\rm SM}}$ is the covariant brane Lagrangian
(that we will identify with the SM Lagrangian), while
the dots indicate all terms involving higher derivatives, the Riemann tensor 
for the induced metric \cite{Giudice:2001av} or the extrinsic curvature. 
By expanding the tension term up to quadratic order in the branons and gravitons we get
\begin{equation}
{\cal L}_{\rm mix}= -\frac{\tau}{2} \left[(\partial_\mu \xi_i) (\partial^\mu \xi^i)+
\kappa(h_\mu^\mu+\xi^i \partial_i h_\mu^\mu+
 2 h_{i\mu } \partial^{\mu} \xi^i) + \kappa^2 B^{\mu\nu\rho\sigma} h_{\mu\nu}h_{\rho\sigma} \right] + \cdots
\end{equation}
which shows a mixing between $\xi$ and $h$ (see appendix B for the definition of the tensor $B$). 
As we will discuss shortly, in order to
consistently compute virtual graviton effects this mixing has to be taken into account.
Notice also that there is a  linear term in $h$, since the massive brane is a source of
gravity. We will comment below about when and how can this term be neglected. The interaction
of gravitons and branons with SM fields is encoded in the covariant  dependence of 
${\cal L}_{\rm SM}$ on the induced metric. At quadratic order we have
\begin{equation}
{\cal L_{\rm SM}} =  L_{\rm SM} + 
 \kappa L^{\mu\nu} \tilde h_{\mu\nu} + 
 \kappa^2 L^{\mu\nu\mu'\nu'} \tilde h_{\mu\nu}\tilde h_{\mu'\nu}+\cdots\qquad
 (T_{\mu\nu}\equiv-2 L_{\mu\nu})
\end{equation}
where the explicit formulae are given in appendix~\ref{vertici}.

\subsection{Gravity and vector bosons} \label{other}

In section \ref{Pure gravity} we have described the gauge fixing procedure for a theory of pure gravity. 
If gauge fields $A_M$ are present, the Lagrangian has both internal and gravitational gauge invariance,
which can be fixed through a delta functional $\delta(F(h,A))$
\begin{equation*}
F(h,A) = [ f_1(h,A),f_2(h,A) ]
\end{equation*}
in the functional integral imposing $f_1(A,h)=0$, $f_2(A,h)=0$.  This is equivalent to 
adding the gauge fixing term
${\cal L}_{\rm GF} = -f_1^2/\xi -f_2^2/2\zeta$ in the Lagrangian
and the Faddeev-Popov determinant in the functional
integral
\begin{equation}
\det \frac{\delta F(h,A)}{\delta\lambda}=\det \begin{pmatrix} 
 \delta f_1/\delta\lambda_1 & \delta f_1/\delta\lambda_2 \\
 \delta f_2/\delta\lambda_1 & \delta f_2/\delta\lambda_2 \end{pmatrix}
\end{equation}
where $\lambda_1$, $\lambda_2$ are the gauge parameters
for diffeomorphisms and internal gauge transformations respectively.
$\delta f_2/\delta \lambda_1$ is generically non zero because vector bosons are `charged' under gravity.
However the graviton field is neutral under charge transformations, so that for a reasonable 
gauge-fixing function
$f_1$ which doesn't involve the vector bosons (in particular for $f_1$ as in eq.~(\ref{eq:GF})) 
the determinant factorizes:
\begin{equation}
\det \frac{\delta F}{\delta\lambda} =
 \det \frac{\delta f_1}{\delta \lambda_1} \cdot \det \frac{\delta f_2}{\delta \lambda_2}
\end{equation}
and the two factors can be exponentiated separately in the usual way.
Notice that  it is convenient to choose a non covariant gauge-fixing $f_2$ for the photons 
in order to avoid  additional couplings with gravitons. (A non-covariant $f_2$ should
not cause any panic: reparametrizations are already broken by the gravitational gauge fixing $f_1$).
We have explicitly checked that the simple gauge fixing
\begin{equation} \label{eq:simpleGF}
{\cal L}_{\rm GF} = -\frac{1}{2\zeta} (\partial_M A_N \eta^{MN})^2
\end{equation}
gives the same results as other more involved choices.

In a theory with vectors that acquire mass $M_v$ through the Higgs
mechanism, as in the Standard Model, the gauge-fixing term will contain the
Goldstone bosons $\phi_G$ field as well. Even with the simple gauge fixing
\begin{equation}
{\cal L}_{\rm GF} = -\frac{1}{2\zeta}  (\partial_M A_N \eta^{MN}-M_v \zeta \phi_G)^2
\end{equation}
there is a cubic vector-Goldstone-graviton interaction:  in a generic metric the gauge fixing does not
fully cancel
the kinetic mixing between the Goldstones and the vector 
($M_v A_M \partial_N \phi_G g^{MN}\sqrt{g}$) present in the Lagrangian.
Such gauge fixing can be easily adapted to vector bosons confined on a brane.

\subsection{Gravity and fermions} \label{fermions}

Finally, we sketch how to extend our analysis to the important case of fermions.
It is well known that GL$(D)$ does not admit spinor representations and in  order to deal with fermions, 
we need some extra structure: the vierbein $E^A_{\,\, M}$ and its inverse $E_A^{\,\, M}$ defined by
\begin{equation}
g_{MN} = \eta_{AB} \, E^{A}_{\,\, M}  E^{B}_{\,\, N} \, \, \qquad  
E^{A}_{\,\,M} E_B^{\,\, M}   = \delta^A_B \; .
\label{vier}
\end{equation}
Where  capital letters from the beginning of the latin alphabet $A,B,C, \dots$ 
denote $D$-dimensional Lorentz  indices.
 The vierbein basis definition introduces an additional gauge symmetry,
besides diffeomorphisms,  due  to the freedom in (\ref{vier}) to rotate  $E$ acting with a 
local SO$(D-1,1)$ transformation. In absence of torsion, the compatibility condition between 
the metric and the connection $\omega$,
allows to express the latter in terms of the vierbein $E$. Then, once the 
vierbein is defined, the introduction of spinors
is rather straightforward (see for instance~\cite{spinors}), 
a collection of the relevant formulae can be found in appendix B. Around a flat background we can parametrize 
the vierbein as $E^A_{\,\, M}=\delta^A_M+\kappa B^A_{\,\,M}$. In terms of the quantum field $B^A_{\,\,M}$
the metric fluctuation is then
%
% In this formalism the gravitation fluctuations around the flat background $\bar e^a_{{}_M}$ are described by
% the quantum vierbein field $e^a_{{}_M}$
% \begin{equation}
% h_{{}_{M N}}   =    \bar{e}^a_{{}_M}  e^b_{{}_N}  + e^a_{{}_M}  \bar{e}^b_{{}_N} +  \kappa \,  
% e^a_{{}_M}  e^b_{{}_N} \, , \qquad \eta_{{}_{M N}}   =  \bar{e}^a_{{}_M} \bar{e}^b_{{}_N}  .
% \end{equation}
% An important observation is that with a suitable gauge choice for the local Lorentz symmetry
% one can trade off the vierbein fluctuation for the metric fluctuation $h$. 
% Given a background metric and the associated background vierbein $\bar{e}$ 
% it is convenient to introduce the tensor
% \begin{equation} 
% B_{{}_{M N}}  =   \bar{e}^a_{{}_M} \,   e^b_{{}_N} \, \eta_{ab} . 
% \end{equation}
% $B$ gives an equivalent parametrization of the  vierbein fluctuation; the metric fluctuation $h$ can be 
% written as
\begin{equation}
h_{{}_{M N}}   =  B_{{}_{M N}}  +  B_{{}_{N M}}    +  \kappa   \,   {B^{{}_{O}}}_{{}_M} B_{{}_{O N}}  .
\label{rel}
\end{equation}
where $B_{M N}=\eta_{MA}B^A_{\,\, N}$ and similarly all indices are raised and lowered
by the Minkowski metric $\eta_{AB}$.
The gravitational action, when expressed in terms of $E$ (or equivalently in terms 
of $B$),  is invariant under the infinitesimal local Lorentz transformation
\begin{equation} 
\delta B^A_{\,\,M} = \kappa^{-1}\Omega^A_B(X) \, \left( \delta^A_M + \kappa B^B_{\,\, M} \right)\, , 
\qquad \Omega_{AB}(X) = - \Omega_{BA}(X)  .
\end{equation}
 A convenient gauge choice is~\cite{des} 
\begin{equation}
B_{{}_{M N}} \, - \,B_{{}_{N M}}  \, = \, 0 ,
\label{gf}
\end{equation}
The great advantage of (\ref{gf}) is that Lorentz ghosts are absent \cite{des} and that it makes possible 
the elimination of the vierbein fields, order by order in $\kappa$,  in favor of the quantum metric $h$ 
\cite{woo}. Indeed, in the gauge (\ref{gf}) one can easily express  $B$ in terms of $h$ by solving 
(\ref{rel}) 
\begin{equation} 
B_{{}_{M N}} = \frac{1}{2} h_{{}_{M N}} - \frac{1}{8} \kappa \,  {h^{{}_{A}}}_{{}_M} h_{{}_{A N}}
+ \, \mbox{O}(\kappa^2).
\label{bfield}
\end{equation}
As a result, even when fermions are present, at the perturbative level, the quantum fluctuations
of the geometry are encoded in $h$ and our formalism can be applied without modifications.

A similar procedure can be applied to fermions living on a $(d-1)$-brane. These are spinors
of SO$(d-1,1)$ and in order to write an invariant Lagrangian one needs the induced
vierbein on the brane, which is now a $d\times d$ matrix $e^a_{\,\,\mu}$. In what follows
we indicate by lower-case latin letters $a,b,c,\dots$ the $d$-dimensional Lorentz indices.
In ref. \cite{sundrum} it was shown how to construct $e^a_{\,\,\mu}$ out of $E^A_{\,\, M}$ and
of the brane immersion $X^M(z)$. Basically it has the form
\begin{equation}
e^a_{\,\,\mu}(z)=R^a_A E^A_{\,\, M}\left (X(z)\right )\partial_\mu X^M(z)
\end{equation}
where $R^a_A$ is a SO$(D-1,1)$ rotation matrix which depends on $E^A_{\,\, M}$ and $X^M(z)$.
Under a SO$(D-1,1)$ rotation $E^A_{\,\, M}\to \Omega(X)^A_B E^B_{\,\, M} $, the induced vierbein
undergoes a SO$(d-1,1)$ rotation $e^a_{\,\, \mu}\to \omega(z)^a_b e^b_{\,\, \mu} $.
By fixing the brane reprametrizations keeping just the branons (as done in the previous section)
and by fixing $D$-dimensional Lorentz transformations as shown in this section, $e^a_{\,\,\mu}$
is written as a function of  $\xi^i(x)$ and  $h_{MN}$. However it is a fairly complicated expression.
Calculations can be simplified by using the local Lorentz symmetry $e^a_{\,\,\mu}
\to \omega^a_b(x) e^b_{\,\,\mu}$ to rotate the induced vierbein to a more convenient form (fermions 
rotate $\psi^\alpha\to \omega^\alpha_\beta \psi^\beta$ by the spinorial representation
$\omega^\alpha_\beta$). Precisely as we did with $E^A_{\,\, M}$ it is useful to rotate $e^a_{\,\,\mu}$
to a symmetric matrix 
\begin{equation}
e^a_{\,\,\mu}=\delta^a_\mu + b^a_{\,\, \mu}\quad\quad \eta_{\nu a}b^a_{\,\,\mu}=\eta_{\mu a}b^a_{\,\,\nu}
\end{equation}
from which by using $g_{\mu\nu}^{\rm ind}=e^a_{\,\,\mu}e^{a\nu}$ and eq.~(\ref{eq:heff}) we  obtain the analogue
of eq.~(\ref{bfield})
\begin{equation}
b_{\mu\nu}=\frac{1}{2}\tilde h_{\mu\nu}-\frac{1}{8}\tilde h^\rho_{\,\,\mu} \tilde h_{\rho\nu}+O(\tilde h^3).
\end{equation}

\section{Loop corrections to  brane observables}\label{toy}
We now have all the ingredients to perform some illustrative computations. We
will focus on the one loop correction to the masses of scalars and vectors
living on the brane.

In order to do a meaningful computation we must employ a regularization that respects 
$D$-dimensional reparametrization invariance.
The result will depend on the choice of the regulator.
The simplest thing could be cutting the loop integrals at $\Lambda$.
Since this is not an invariant regulator, we would get a meaningless result that also 
depends on the choice of the loop integration momentum. 
A better possibility consists in dividing all graviton propagators by some power of 
$(1-p^2/\Lambda^2)$.
It is possible to obtain this Pauli-Villars (PV) regulator in a covariant way
by adding to the action suitable  higher derivative reparametrization invariant
terms involving just the metric.
We will instead employ the standard extension of dimensional regularization to the case in which
both continuum and discrete momentum are involved (see the appendices for details).
Of course by this method we are only sensitive to the ``physical'' logarithmic divergences, 
while all power 
divergences are automatically removed. Nonetheless from our results it will be clear that by
choosing a regulator sensitive to power divergences (like PV) for the sum over KK
we would still not have the $\Lambda R$ terms of ref.s~\cite{KKgr2,EWPOgr1,EWPOgr2,Bij}
in physical quantities.

\subsection{Brane in a torus}

Consider now  the one-loop graviton correction to the {\em pole} mass $m_0$ 
of a minimally coupled  scalar
living on a {\em straight} brane located at the point $y^i=0$ of a torus $T^\delta$.
As we explained in 
section \ref{brane}, in the regime of validity of  EFT ($E<1/r_G$)
the brane tension can be treated as a perturbation in the gravitational dynamics. 
Therefore it makes sense to expand the corrections to our observables in a power
series in $\tau$. Let us focus on the lowest order effects, $i.e.$ those
that go like $\tau ^0$ \footnote{Notice that there are also corrections from pure branon
exchange, which go like inverse powers of $\tau$ and which persist when gravity
is turned off. The lowest, physically meaningful  correction of this type to the scalar mass
comes at two loops and goes like $\delta m_0/m_0\sim m_0^8/\tau^2$. The $1/\tau$ effects
can be bigger than the gravitational ones we study, but they are physically independent~\cite{branoni}.
Thus it makes sense to  focus only on the latter.}. The diagrams that contribute at order
$\tau^0$ are shown in fig.~1. Notice that diagrams (d) and (e) also involve branons:
in these diagrams the $\tau^{-1}$ from branon propagation is compensated by
the $\tau^1$ in the graviton-branon mixing insertion. 
Notice also that the tadpole diagram (c) gives no contribution. Due to momentum conservation 
in the extra dimensions (valid at zeroth order in $\tau$) only the zero modes
mediate this tadpole, these are the 4d graviton and the radion. Whatever mechanism 
stabilizes the radion giving also a vanishing effective 4d cosmological
constant generates a tadpole that cancels (c) exactly at the minimum of
the radion potential. Of course exact cancellation of the 4d cosmological constant requires
the usual fine tuning.
Now, the genuine graviton diagrams (a) and (b) give a correction
\begin{equation} \label{eq:circle1}
\delta m^2_0 (\hbox{a}+\hbox{b}) = \frac{i}{\bar{M}_d^{d-2}}\sum_n
 \int_k \langle{\phi(-p)}|-2L^{\mu\nu} L^{\rho\sigma}+4i L^{\mu\nu\rho\sigma}|\phi(p)\rangle 
(\Tcontr{0.2em}{2.5ex}{1ex}{1.6em}h_{\mu\nu}^{(n)} h_{\rho\sigma}^{(-n)})
\end{equation}
where $p^2=m^2_0$ is the squared momentum of the on-shell scalars ($m_0$ is the tree level mass).
\begin{figure}
\centering
\epsfig{file=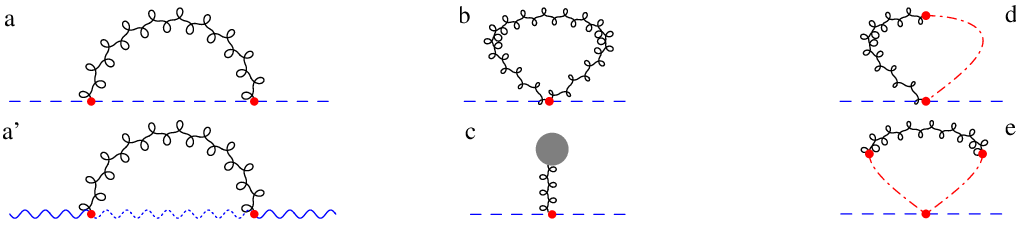,width=1.1\linewidth}
\caption{\em
One-loop gravitational corrections to the pole mass of a scalar on a brane from gravitons
(diagrams a,b,c) and from graviton/branon mixing (diagrams d,e) at zeroth order in the brane tension.
The mass of a vector particle on a brane also gets corrections from vector-Goldstone-graviton 
vertices (diagram a$'$). Gravitons (branons, scalars, vectors, Goldstones) are drawn as pig-tail
(dot-dashed, dashed, wavy, dashed-wavy) lines.}
\label{fig:grafici}
\end{figure}
The branon contributions (d) and (e) give
\begin{equation} \label{eq:circle2}
\delta m^2_0(\hbox{d}+\hbox{e})=\frac{i}{\bar{M}_d^{d-2}} \langle{\phi(-p)}|T_{\mu\mu}|\phi(p)\rangle   
 \sum_{n}\int_k  F_n(k)
\end{equation}
where $\langle{\phi(-p)}|T_{\mu\mu}|\phi(p)\rangle = 2m^2_0$ and $F_n(k)$ represents
the contribution of the loop in fig.~(1d,e).
The sum of all  contributions in the interpolating $\xi$ gauge defined in section 2 gives
\begin{equation} \label{explicit}
\begin{split} \delta m_0^2 &= 
         \frac{m_0^2}{{32 \bar M_d^{d-2} \pi^2 d(d+\delta-2)}} \sum_n \Big\{ 
         f(\xi,d,\delta) A_0(\Mn^2) - d(d-2) (\delta-4) A_0(m_0^2) + \\
        &  4d\big[ (d-2) \Mn^2  -2 (d+\delta-3)m_0^2 \big] B_0(m_0^2,\Mn^2,m_0^2) +
         d\delta(d-2) \Mn^2 B_1(m_0^2,\Mn^2,m_0^2) \Big\} \\[0.3cm]
\end{split} 
\end{equation}
\begin{equation*}
\begin{split}
f(\xi,d,\delta) =
 & 4 (2\xi-1)^{d/2-1} \big[2-\delta(2\xi-1)(\xi-1)-\xi(10-3d+3\xi(d-4))\big]+\\
 & 2\xi^{d/2} \big[3d\delta+d\xi(\delta-2)+d^2 (\xi+1)+4-12\xi+2\delta(\xi+\delta-4) \big]+ \\
 & 2(d-2) \big[-\delta+2d(d+\delta-2)\big]
\end{split}
\end{equation*}
where the Passarino-Veltman functions $A_0$, $B_0$, $B_1$ are defined in appendix~D, where we describe
how the cutoff-independent contribution can be extracted.
The mass correction is multiplicative as expected for a minimally coupled scalar:
for vanishing tree level mass the scalar is derivatively coupled to  gravity.
Although in this expression all the terms enhanced
by $1/\Mn^4$ found in~\cite{KKgr2,EWPOgr1,EWPOgr2,Bij}
cancel out mode by mode, we do {\em not}
 obtain a gauge-independent result. However the
$\xi$ dependent term in $\delta m_0^2/m_0^2$, only depends on $M_D$ and $R$ (and
the UV cut-off $\Lambda_{\rm UV}$ if dimensional regularization is abandoned) but not on $m_0^2$ itself.  
So it looks like a universal effect. Indeed one finds the same gauge dependent
piece in the correction to the mass of a vector particle. In short 
the correction to the pole mass of a spin $s=\{0,1\}$ particle 
on the brane can be written as
\begin{equation}\label{eq:dmbrane}
\delta m^2_s = 2 m_s^2 G(M_DR) + m_s^2 \Delta_s(m_s,\bar{\mu} ,R, M_D)
\end{equation}
where $G$ is the only gauge dependent factor,
while $\Delta_0,\Delta_1$ are gauge-independent 
($\bar{\mu}$  is the renormalization scale).
Explicit expressions for these functions can be found in appendix~\ref{risultati}.
Similarly we find that for a localized photon the gravitational correction to the 
electric charge $e$ has a  gauge-dependent
factor equal to $[e]G$, where
$[e]=2-d/2$ denotes the dimension of the electric charge in $d$ dimensions. The moral 
of these results is that
 the gauge dependence  can be reabsorbed by changing the normalization 
of the graviton field $g_{MN}$.
In more physical terms, gauge dependent terms amount to a change of the mass unit:
all the dimensionless quantities that we have computed (like $m_0/m_1$) are gauge independent.
In the presence of gravity only dimensionless quantities are real observables, as they are 
invariant  under rescaling of the metric\footnote{The gauge dependence of pole masses
in quantum gravity was already found and discussed in ref.~\cite{tomaras}. In the next
section we will give a simple geometrical explanation.}.
As a simple further check we have also
computed the  corrections to the masses of bulk particles. In  particular we have 
focused on the $n=0$  modes of fields with
spin $s=\{0,1\}$ and  bulk mass $\{m'_0, m'_1\}$. Again, we obtain
\begin{equation}\label{eq:dmbulk}
\delta m^{\prime 2}_s = 2m_s^{\prime 2} G + m_s^{\prime 2} \Delta'_s.
\end{equation}
where the gauge dependent part is the same as for brane modes but the physically
meaningful piece $\Delta'_s$ is, as expected, different. Notice that bulk particles
 do not couple directly to branons, so that there is no analogue of diagrams (d) and (e)
for them. On the other hand the bulk modes couple directly 
to the $h_{i\mu}$ and $h_{ij}$ pieces of the graviton field,
which was not the case for the brane modes.  In view of these differences,
the fact that the gauge depended piece is always the same is a rather
non trivial check. 

A concluding remark on the $1/\Mn^4$ terms found in~\cite{KKgr2,EWPOgr1,EWPOgr2,Bij}
 is in order. These  terms come only from diagram
(a), so that the branons play no role in the cancellation of these effects in physical quantities. 
Furthermore, it is clear that  terms of this type
could not be physical, as they cannot arise in the $\xi=1$ gauge. However in
gauge dependent quantities they can appear. In the appendix we give the expression
of the scalar self-energy at the off-shell point $p_{\rm ext}=0$, where these unphysical
effects are indeed present.
Notice that if they appeared in physical quantities
there would   really be
an enhancement of the result by some power of the
radius $R$ (IR divergences cannot be thrown away!). 

So far we have only considered observables
that do not depend at tree level on the size $R$ of the compact extra dimension.
A gauge invariant result is obtained in a slightly more complicated way when one considers
observables like $m_0/M_{\rm Pl}$ or the ratio between pole masses of different KK excitations.
The reason is that $R$ itself is gauge dependent: a discussion of this issue,
including a geometrical explanation of this statement, is presented in subsection~\ref{general}.

\subsection{Gauge independence of physics and geometry\label{general}}

In this section we want to extend our discussion of gauge invariance to generic observables
that depend on the size $R$ of the extra dimensions.
To be concrete we will compare two gauge
choices,  unitary ($U$) and de Donder ($DD$). 
The compact manifold is assumed to be a $\delta$-torus.

 The previous results
could be restated as follows: in order to get the same physics
in the $U$ and $DD$ gauges, all the tree level mass parameters $m_U$ and $m_{DD}$ 
in the two gauges should be related by 
\begin{equation}
m_U^2=m_{DD}^2\big[1+[m^2](G_{U}-G_{DD})\big]=
m_{DD}^2\lambda_1\qquad [m^2]=2,
\end{equation}
where the $G$'s are the universal quantities given in appendix~C. 
Similar relations hold for parameters with different mass dimensions.
This
is equivalent to taking as background metrics $\lambda_1\eta_{MN}$
and  $\eta_{MN}$ in respectively the $U$ and $DD$ gauge, but keeping the same tree level
mass parameters ($i.e.$ $m_{DD}$). This is easily seen because
the tree level Klein-Gordon operator in $U$ gauge is $\lambda_1^{-1}\eta^{MN}
\partial_M\partial_N+m_{DD}^2$.

\begin{figure}
\centering
\epsfig{file=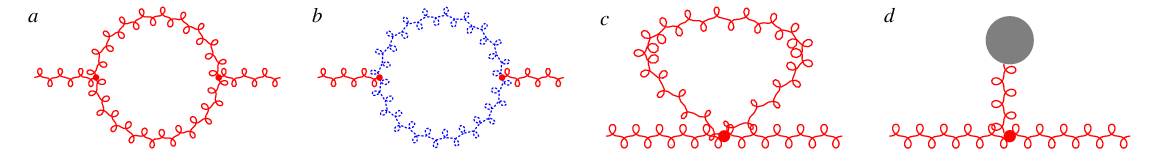,width=0.9\linewidth}
\caption{\em
The corrections to the Planck mass is obtained by combining corrections to the graviton propagator
(we show the relative Feynman diagrams. Diagram b contains a ghost loop) with corrections to the graviton/matter vertex
and with corrections to the matter propagator.}
\label{fig:calcoloni}
\end{figure}

This argument is basically correct, but not completely. The point is that
since the space is not isotropic (there are $\delta$ compact directions)
the metric rescaling factor $\lambda$ does not have to be 
the same for all directions. Then,
compatibly with the symmetries of the system, we expect in general
the backgrounds
\begin{equation}
g_{MN}^{DD}=\begin{pmatrix}
\eta_{\mu\nu}&0\\ 0&\eta_{ij}\\ \end{pmatrix}\quad\quad
g_{MN}^U=\begin{pmatrix}\lambda_1\eta_{\mu\nu}&0\\ 0&\lambda_2\eta_{ij}\\
\end{pmatrix}
\end{equation}
with $\lambda_1\not = \lambda_2$. These relative backgrounds have to
be chosen in order to get the same results in the two gauges. Notice
that we keep the same periodicity $y_i\sim y_i+2\pi R$ on the torus.
Then, at tree level, the proper length of the period of the torus
is rescaled by a factor $\sqrt{\lambda_2}$ in the unitary gauge. 
In the same gauge the mass shell condition for the mode $\{n_i\}$ is 
\begin{equation}
(\eta^{\mu\nu}\partial_\mu\partial_\nu+\frac{\lambda_1}{\lambda_2}\frac{n^2}{R^2}+\lambda_1
m^2)\phi_n=0
\end{equation}
so that $R$ as defined through KK masses is rescaled by 
$\sqrt{\lambda_2/\lambda_1}$ at tree level, and not by $\sqrt {\lambda_2}$.
Finally the tree level $d$-dimensional Newton constant $G_N=1/(M_D^{d+\delta-2}
R^\delta)$ in the $U$ gauge is given by $G_N^U=G_N\lambda_2^{-\delta/2}
\lambda_1^{1-d/2}$. By writing $\lambda_i=1+c_i$, at lowest order in the $c_i$
we then have the tree level relations
\begin{align} 
m_U  =& m_{DD}(1+\frac{c_1}{2}) \nonumber \\
R_U  =& R_{DD}(1+\frac{c_2}{2}-\frac{c_1}{2})  \label{sys:rescalings}  \\  
(G_N)_U =& (G_N)_{DD}(1+(1-\frac{d}{2})c_1-\frac{\delta}{2}c_2)\nonumber
\end{align}
where the radius is here defined through the KK masses. In our calculations
so far we only considered the masses of brane modes or bulk zero modes, which 
do not depend on the radius at tree level. This is why one universal rescaling 
$\lambda_1$ was enough to eliminate spurious gauge effects. By considering
the quantum corrections to KK masses or to the $d$-dimensional Newton constant
one finds extra gauge dependence. 
However we have checked by explicit calculations that
it can all be eliminated consistently with eq.~(\ref{sys:rescalings}).
The corrections to KK masses  represent just a direct generalization of the
computation of the previous section. On the other hand, the Newton constant requires to compute
also the correction to the graviton-matter vertex and to the graviton propagator.
Few relevant Feynman diagrams are shown in fig.~\ref{fig:calcoloni}.
This is a lengthy computation\footnote{For example
in the unitary gauge the diagram~\ref{fig:calcoloni}a is obtained by summing 2.588.740 terms.
All computations in this work have been done with Mathematica~\cite{Mathematica}.}
 upon which the gauge dependence in eq.~(\ref{sys:rescalings})
is a non-trivial check. At one loop, we find
\begin{equation}
\begin{split}
\lambda_1 &= 1+ \frac{d^2 +d(2\delta-1)+\delta(\delta-3)}{8 \pi^2 \bar M_d^{d-2} d(d+\delta-2)}
 \sum_n A_0(\Mn^2) \\
\lambda_2 &= 1+ \frac{d}{8 \pi^2 \bar M_d^{d-2}} \sum_n A_0(\Mn^2)
\end{split}
\label{cici}
\end{equation}
It is interesting to re-derive the quantities $\lambda_1,\lambda_2$ in a purely
geometrical way. For instance $\lambda_2$ is fixed by the coordinate
independent proper period of the torus. We can easily show this for the case 
$\delta=1$, $d=4$ (the latter choice being made just to simplify the notation).
In order to do so we must  (arbitrarily) pick a path around the compact dimension,
and make sure that working in different gauges the path is kept unchanged.
It is convenient to simply pick the path ${\cal P}$ defined in unitary
coordinates by $x^\mu=0, y^5=\tau$  with $\tau$ going from $0$ to $2\pi R$. 
Actually any path in the {\it family} $x^\mu={\rm const}, y^5=\tau$ ($\tau=[0,2\pi R]$)
would give the same result as it is  equivalent by translation invariance 
of the background\footnote{This is
an important point since by construction the unitary coordinate frame, defined
by the request that  $g_{55}$ and $g_{\mu 5}$
be independent of $y^5$, is truly a {\it family} of gauges. This is because 
the unitary form of the metric is preserved by
the ``zero mode'' coordinate changes $x^\mu\to x^\mu+\epsilon^\mu(x)$, 
$y^5\to y^5+\epsilon^5(x)$ (corresponding to $4$-dimensional diffeomorfisms and to the circle isometry). Then
since our path ${\cal P}$ is defined in a family of gauges it truly designates a
family of paths. It is manifest that this family corresponds to the paths
related to ${\cal P}$ by translation in $x$. They all have the same length.}.
Notice also that in a general coordinate choice, $x^\mu$ is not constant
along ${\cal P}$.  In an arbitrary coordinate system the definition $X^{M}(\tau)$
of ${\cal P}$ 
depends also on the metric itself. It is easy to work out this dependence.
The quantity
\begin{equation}
L=\bigg\langle\int_{\cal P}\sqrt{g_{MN}\dot X^M\dot X^N}d\tau \bigg\rangle
\end{equation}
must be gauge independent, being the expectation value of a gauge invariant
operator. Comparing 
the calculation of $L$ in the $U$ and $DD$ gauges, we get at 1-loop order 
\begin{equation*}
\begin{split}
L = &2\pi R+\frac{c_2}{2}-\frac{\kappa^2}{8} \langle  h_{55}^{(0)}h_{55}^{(0)} \rangle _U 
 = 2\pi R-\frac{\kappa^2}{8}\langle  h_{55}^{(0)}h_{55}^{(0)} \rangle _{DD} \\ &+
 \frac{\kappa^2}{2} \sum_{n\not = 0} \bigg[ 
 \eta^{\mu\nu}  \langle h_{5\mu}^{(n)}h_{5\nu}^{(-n)} \rangle _{DD} + \frac{1}{4} 
 \Big(\frac{\eta^{\mu\nu} \langle\partial_\mu h_{55}^{(n)}\partial_\nu h_{55}^{(-n)}\rangle _{DD}}{\Mn^2}-
  \langle h_{55}^{(n)} h_{55}^{(-n)}\rangle _{DD} \Big) \bigg].
\end{split}
\end{equation*}
In the unitary gauge only the scalar zero mode (radion) $h_{55}$
contributes to $L$, while in the $DD$ gauge extra contributions from KK graviphotons and graviscalars show up
(the latter is zero in dimensional regularization).
However in the $U$ gauge there is the tree level term $c_2$. Notice that in both
gauges the gravitational field $h$ is defined to have no tadpoles.
This equation
fixes $c_2$ and the result agrees with what found for the physical
parameters.

Finally we can fix $c_1$ by considering the volume element in the 
two gauges
\begin{equation}
\bigg\langle  \int_{T^{\delta}}\sqrt{g}~d^\delta y ~d^d x \bigg\rangle _U=
\bigg\langle  \int_{T^{\delta}}\sqrt{g}~d^\delta y ~d^d x \bigg\rangle _{DD}
\label{volumes}
\end{equation}
where we have fully integrated on the torus  $T^\delta$. This equality
ensures that the non-compact coordinates $x$ represent
the same physical distance in the two gauges. The mass of a
particle, as defined by the $x$ dependence of the propagator,
has then to be the same in the two gauges. Taking the background into account,  eq.~(\ref{volumes})
reads at 1-loop
\begin{equation}
(2\pi R)^\delta\left [1+\frac{\delta c_2}{2}+\frac{d c_1}{2}+
\langle  \hat O \rangle _U\right]=(2\pi R)^\delta\left [1+
\langle  \hat O \rangle _{DD}\right ]
\end{equation}
where
\begin{equation}
\hat O= \frac{\kappa^2}{8}\sum_{n}(\eta^{MN}\eta^{RS}-2\eta^{MR}\eta^{NS})
 h^{(n)}_{MN} h^{(-n)}_{RS}.
\end{equation}
Using the previous result for $c_2$ we determine $c_1$ in agreement with eq.~(\ref{cici}).

\subsection{Higher orders in the brane tension}

\begin{figure}
\centering
\epsfig{file=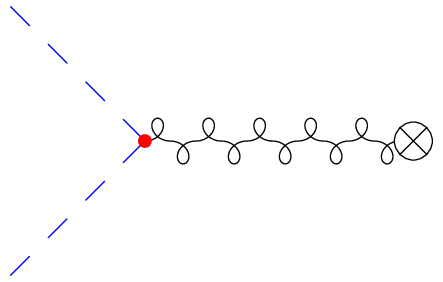,width=0.2\linewidth}
\caption{\em A diagram of order $\tau^1$.}
\label{selffield}
\end{figure}

We have only studied  the terms of zeroth order in the tension
$\tau$, but things should work out in a similar way order by order in $\tau$.
These higher order effects come not only from branon insertion in the diagrams
of fig.~(1a) and (1b) but also from extra tadpoles. Indeed at order $\tau$ there is
already at the tree level the tadpole of fig.~(\ref{selffield}).
It corresponds to the brane self gravitational field. 
Of course if we treat the brane as a thin object this field
is infinite at the brane itself. This is a UV divergence which is eliminated
by adding the suitable counterterms (amounting to a renormalization of
the unit length on the brane). Applying our regulator (see appendix~\ref{integrals}) 
we get from fig.~(\ref{selffield}), for $d=4$
\begin{equation} \label{eq:taudiag}
\delta m_0^2= m_0^2\frac{\mu^4}{M_4^2 R^{\delta}} \frac{8(\delta-2)}{(\delta+2)} \sum_n \frac{1}{\Mn^2}=
m_0^2 \frac{\mu^4}{M_D^{2+\delta} R^{\delta-2}} \frac{8(\delta-2)}{(\delta+2)} \, {\cal I}_1
\end{equation}
where ${\cal I}_1$ is a constant defined in appendix~\ref{integrals}.
The  $R$ dependence is insensitive to  the UV cut-off: it measures the deformation of
the brane self field due to the finite volume, so it is a well defined quantity 
in the EFT approach.  Notice also that fig.~(\ref{selffield}) is a brane-to-brane
exchange of a bulk graviton like those considered in various phenomenological studies
\cite{GRW,virtual}. At one loop fig.~(\ref{selffield}) is dressed into extra tadpole diagrams:
we expect that inclusion of these tadpoles will be essential to get gauge independent
results at linear order in $\tau$.

\subsection{Brane in an orbifold}

In section~\ref{brane} we explained that, for a brane
living on a smooth space,
the branons have to
be kept in order to preserve general covariance. Then
we have explicitly shown that the branons are needed to restore reparametrization
gauge independence of quantum gravity corrections. In this section we show
an example of how things work for a brane stuck at a fixed point of
an orbifold. 
Now the brane cannot move, {i.e.} there is no branon degree of freedom. 
But at the same time the group of diffeomorfisms is also changed. Indeed when
dealing with fixed points it is even superfluous to talk about a brane:
at these points we can localize degrees of freedom and interactions respecting
the orbifold reparametrization invariance. For simplicity we will 
consider  the simplest case of a brane in
$\mathbb{R}^d \times S^1/\mathbb{Z}_2$. 
The space $S^1/\mathbb{Z}_2$ is a line segment,
obtained identifying points in a circle of radius $R$
according to the $\mathbb{Z}_2$ reflection:
\begin{equation}
y \sim 2\pi R -y; \qquad y\in [0,\,2\pi R]
\end{equation}
which has $0,\pi R$ as fixed points.
The invariance of the line element $ds^2$ under $\mathbb{Z}_2$ implies that under the orbifold
reflection
the metric components $h_{\mu\nu}$, $h_{ij}$ and the ghost field $\eta_\mu$ are even,
while $h_{\mu i}$ and $\eta_i$ are odd.
A generic field $f(x,y)$ can be Fourier decomposed according to its parity:
\begin{equation}
\begin{gathered}
f(x,y) = \sum_{n=0}^{+\infty} f^{(n)}(x) \, \Psi_n(y), \qquad
\Psi_n(y) = \begin{cases} 
 a_n \cos(ny/R)  &  \quad \text{even} \\[0.1cm]
 b_n \sin (ny/R) &  \quad \text{odd} \end{cases} \\[0.3cm]
a_0 = \frac{1}{\sqrt{2\pi R}}, \quad b_0=0  \qquad \qquad  
a_n =b_n= \frac{1}{\sqrt{\pi R}} \quad n\not=0.
\end{gathered}
\end{equation}
Odd fields do not have a zero mode.
We can use the same gauge fixing for reparametrization invariance as before.

The group of diffeomorfisms on the orbifold is defined by the transformations
$x^\mu\to f^\mu(x,y)$, $y\to f^5(x,y)$
with $f^\mu$ even and $f^5$ odd under orbifold reflection (both $f^\mu$ and $f^5$ have
period $2\pi R$). Notice that the  boundaries $y=0$ and $y=\pi R$ 
are left fixed. A brane at $y=0$ remains a brane at $y=0$ in all reference frames. 
Even if we do not let the brane fluctuate we still obtain consistent results. On the other
hand for a brane at a generic $y\not =0,\pi R$, its position depends on the
reference frame and we are forced to let it fluctuate. What is special about
the fixed points is that 
we have thrown away enough gauge degrees of freedom ($g_{\mu 5}(y=0,\pi R)=0$)
that we can live without branons. Let us see this explicitly.

The computation of the gravitational corrections in the orbifold geometry is similar 
to the previous ones, but with some important differences.
Consider the case of a brane sitting at a generic point $y$. 
Contrary to the circle case, the $y$ dependence in the coupling 
between matter on the brane and gravity does not cancel out in physical amplitudes; 
for instance, the cross section
for the production of an individual  KK graviton mode is proportional to $\cos ^2(ny/R)$.
It is not a surprise that this factor depends on $y$: $S^1/\mathbb{Z}_2$ is 
not an homogeneous space.
Similarly, the branon contribution in graviton loops gets multiplied by a factor $\sin ^2(ny/R)$,
showing that their presence is not necessary when $y= 0$.
However, having altered by a $y$-dependent factor the relative weight between graviton 
and branon effects, we apparently no longer get a gauge invariant result.
We now show that we must take into account a new type of graviton tadpoles
that were absent on a homogenous space.

The conservation law of fifth dimensional momentum is altered since
some of the harmonics are projected out by the $\mathbb{Z}_2$ symmetry.
In a vertex with three lines carrying momenta 
$n_i\geq 0$, $i=1,2,3$ along the fifth dimension it reads $n_1 \pm n_2 \pm n_3 =0$.
The propagator of matter on the brane is corrected by new tadpole diagrams, like (1c), 
but with non zero extra-dimensional
momentum $2n$ on the tadpole graviton line.
The blob in fig.~(1c) can be either a graviton loop or a gravitational ghost loop.
Notice that, while tadpoles with a zero momentum internal line ($n=0$) are assumed to be exactly
canceled by a suitable stabilization mechanism, the same type of diagrams with non zero $n$ must
be taken into account and they are crucial to recover gauge invariance for brane observables.
Let us focus for instance on the mass correction $\delta m_0^2$ for a scalar on a brane
at a generic $y$. Notice that for $y\not=0,\pi R$ the brane is free to move, so branons
must be kept. 
We get
\begin{equation}
\begin{split}
\delta m_0^2 = \sum_n \big[ &2\cos^2 (n y/R)\, F^{(n)} (\text{gravitons}) + 
 2\sin^2 (n y/R)\, F^{(n)}(\text{branons}) + \\ 
 &\cos (2n y/R) \, F^{(n)}(\text{tadpole}) \big] = m_0^2\left(2 G(M_DR)+
\tilde \Delta_{0}(m_0,y,R,M_D,\bar\mu) \right )
\end{split}
\end{equation}
The contribution  from the $n$th KK mode in the graviton diagrams of
fig.~(1a,b) is exactly the same as in the torus, except for an overall factor $2\cos^2 (n y/R)$
coming from the  graviton wave function.
The contribution $F^{(n)}(\text{branons})$ from diagrams in 
fig.~(1d,e) gets instead an overall $y$-dependent factor $2\sin^2 (n y/R)$.
The final result for $\delta m_0^2$ has the same structure of eq.~(\ref{eq:dmbrane}), but now
the gauge invariant piece $\tilde \Delta_{0}$ is a function of $y$.
Finally, the tadpole contribution comes in the right way to cancel mode by mode the $y$-dependence 
in the gauge variant term $G$.
This is consistent with the mass correction for the zero mode of a scalar propagating in the bulk,
which has the same form as in the torus case (see eq.~(\ref{eq:dmbulk})), 
and it represents a non trivial check on the result.
In the special case of a brane sitting at the fixed points $0,\pi R$ the branon contribution
vanishes and gauge invariance of the pole mass is met just through tadpole diagrams.

As a final remark, we notice that because of the modified momentum conservation law in the orbifold,
the zero mode of a bulk field mixes with its KK excitations at one loop level. 
The relevant diagrams are those in fig.s~(1a,b,c) with discrete momentum $n$ in the internal loop and $0$, $2n$ in the external legs. 
This effect however is relevant
only at order $\kappa^4$ and can be safely neglected.

\section{Phenomenology}\label{EWPO}

In the previous sections we have explained how to consistently compute
quantum gravity corrections using an effective field theory (EFT). 
A possible physical application  is the computation of graviton loop corrections to 
electroweak precision
observables (EWPO) and to the anomalous moment of the muon
in brane models with large extra dimensions and a TeV-scale $D$-dimensional Planck mass.
Unfortunately our knowledge of the low energy effective theory of gravity only
allows to reliably compute corrections of little phenomenological interest.
Basically, the EFT allows
to compute those contributions that are saturated in the infrared, i.e.\ at the
scale of the relevant external momenta. For instance, the calculable corrections
to EWPO go like $(M_Z/M_D)^{2+\delta}$
(or  $(M_Z/M_D)^{2+\delta}\ln M_Z$) by simple dimensional analysis.
These effects go to zero very quickly when $M_D$ is raised, becoming negligible already 
for $M_D$  below a TeV. On the other hand, the contributions from the region of
large virtual loop momenta  gives in principle a much larger effect. However,
being saturated in the UV region, where we do not control the EFT, these contributions
are not calculable. 
This problem already affects tree level virtual graviton effects.

We can however {\em estimate\/} graviton effects by introducing an explicit UV cutoff $\Lambda$.
The corrections to EWPO will scale like
$M_Z^2\Lambda^\delta/M_D^{2+\delta}$.
The unknown physical cutoff could perhaps be produced by string theory, 
or could be related to the inverse brane
width or even to just the brane tension~\cite{bando}.
Since we do not know we must keep $\Lambda$ as a phenomenological parameter
and discuss its physical meaning and plausible value.

Virtual graviton corrections (even at tree level)
cannot be computed from Einstein gravity as much as
electroweak quantum corrections  cannot be computed from  Fermi theory.
In the latter case the complete theory is known and perturbative: by
comparing to the full theory one sees 
that correct estimates are obtained by cutting off power divergent four-fermion loops at a
``small'' scale $\Lambda \sim M_W \approx g G_{\rm F}^{-1/2}$
rather than at the larger  $\Lambda \approx G_{\rm F}^{-1/2}$.
At least at a qualitative level, the gravitational $\Lambda$ can be given a similar  physical meaning.

\subsection{Strong vs weak gravity: NDA estimates}

Therefore we first identify the value $\Lambda_S$ of $\Lambda$ that corresponds to strongly coupled quantum gravity\footnote{
In the context of string theory this
corresponds to a situation where the string coupling is essentially at the self
dual point.}.
This can be done by adapting to our case the naive dimensional analysis (NDA)
technique developed to estimate pion interactions below the QCD scale \cite{nda}
(NDA has already been applied to brane models~\cite{luty}).
NDA allows to estimate the size of the effects from
a strongly coupled theory up to coefficients of order 1 but including all the
geometric dependence on powers of $\pi$. By applying NDA, we estimate 
\begin{equation}
\Lambda_S^{2+\delta}\approx \pi^{2-\delta/2}\Gamma(2+\delta/2) M_D^{2+\delta}
\end{equation}
In the range of interesting $\delta$, $\Lambda_S$ is not much larger than $M_D$. 

We first discuss the particular case $\Lambda=\Lambda_S$: 
diagrams with any number of graviton lines give comparable contributions,
and NDA allows to estimate their size.
Tree level exchange of gravitons generates the effective dimension 8 operator
${\cal T}\equiv T_{\mu\nu}^2-T_{\mu\mu}^2/(\delta+2)$~\cite{GRW,KKgr,KKgr2}.
Its coefficient in the effective Lagrangian is divergent, and
NDA estimates it to be $\approx \pi^2/\Lambda_S^4$. 
This operator is however not the most important in low energy phenomenology, because at loop level
gravitons generate dimension 6 four fermion operators
with coefficient $\approx \pi^2/\Lambda_S^2$. On the other hand
the operator  $W_{\mu\nu}^a B^{\mu\nu} H^\dagger \tau^a H$ is generated with coefficient 
$\approx g_2g_1/\Lambda_S^2$ with no $\pi^2$ enhancement. 
This property is shared by other operators that require the exchange
of virtual gauge bosons. This is because we are assuming that the weak gauge couplings 
remain weak up to the cutoff.

By drawing a few Feynman graphs one can see that tree level exchange of gravitons 
(and therefore the operator ${\cal T}$) does not affect precision observables at the  $Z$-resonance. 
Moreover the four fermion operators induced by double graviton exchange are of neutral current type,
so they do not directly affect $\mu$ decay and are therefore not constrained by high precision data.
$\mu$-decay is affected by one loop diagrams with a $W$ and a graviton: their coefficient
is only $\approx g_2^2/\Lambda_S^2$. 

Indeed by a simple analysis one finds  that all dimension six
operators that affect EWPO have a coefficient 
$\approx g^2_{2}/\Lambda_S^2\approx 1/\Lambda_S^2$. 
As shown in~\cite{NRO} EWPO set a bound $\Lambda_S >(5\div 10)\TeV$
on a generic set of dimension 6 operators that conserve baryon, lepton and flavor numbers and CP.
This bound seems rather strong when compared to the sensitivity to direct graviton emission expected 
at the next colliders.
Furthermore since $m_{\rm top}\approx 175\GeV$ a real solution of the hierarchy problem should cutoff
the quadratically divergent top correction to the Higgs mass
at a much lower value of $\Lambda\approx 300\GeV$.
Our assumption $\Lambda = \Lambda_S$ corresponds however to one of the most constrained scenarios:
LEP data strongly disfavor new strongly coupled physics in the electroweak sector.
The situation becomes worse if we assume that also the gauge couplings get strong at $\Lambda=\Lambda_S$.
\begin{figure}
\centering
\epsfig{file=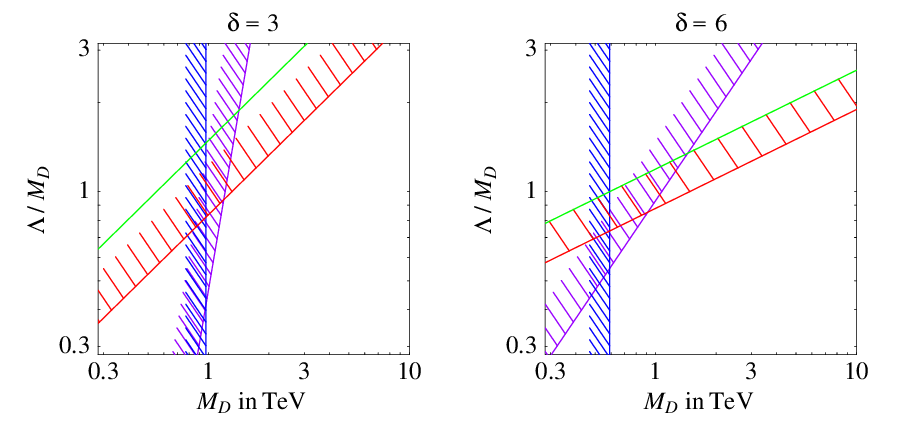,width=0.9\linewidth} 
\caption{\em Compilation of collider bounds on graviton phenomenology in the plane $(M_D,\Lambda/M_D)$ for
$\delta=3$ (left)  and $\delta=6$ (right).\label{fig:vincoli}}
\end{figure}

In order to obtain a more acceptable phenomenology one can assume that the UV cut-off $\Lambda$ happens
to be smaller than $\Lambda_S$, so that gravity does not become strong
and dominant graviton corrections to EWPO are dominated by one loop diagrams
(presumably a complete theory will not contain only gravitons).
In the next subsections we `compute' the graviton corrections to the
electroweak observables (expressed in terms of the $\epsilon_1,\epsilon_2,\epsilon_3$ 
parameters~\cite{epsilon}) and to the anomalous magnetic moment of the muon.
In agreement with NDA estimates, the final result is of the form
\begin{equation*}
\delta\epsilon_i \approx \frac{M_Z^2}{\Lambda_S^2}\bigg(\frac{ \Lambda}{\Lambda_S}\bigg)^\delta,\qquad
\delta a_\mu \approx \frac{m_\mu^2}{\Lambda_S^2}\bigg(\frac{ \Lambda}{\Lambda_S}\bigg)^\delta
\end{equation*}
where the factors of order one depend on the choice of cutoff.
Not knowing which is the physical cutoff,
we use dimensional regularization:
with this choice loop integrals do not give powers of $\Lambda$.
However, since we are considering a higher dimensional theory,
powers of $\Lambda$  arise  from divergent sums over the
KK levels of the gravitons.
A different choice of the cutoff would give different results.

In fig.~\ref{fig:vincoli} we summarize the present situation of collider graviton phenomenology
by collecting the various bounds in the plane $(M_D,\Lambda/M_D)$:
\begin{itemize}
\item The vertical bound comes from emission of real gravitons~\cite{GRW} at LEP2 and Tevatron~\cite{vincoli}.
It does not depend on $\Lambda$ (as long as the energy of the collider is less than $\Lambda$) because it
is the only bound on really computable effects.

\item Virtual exchange of gravitons at tree level generates the operator ${\cal T}$.
Its coefficient depends on the cutoff $\Lambda$ so that it cannot be computed from 
the low energy EFT
(in the literature there exists a variety of estimates~\cite{GRW,KKgr2,virtual}, 
freely dubbed ``formalisms'', 
and a corresponding variety of experimental bounds \cite{vincoli}).
The coefficient can be estimated to be
$\approx \pi^2\delta(\delta+2){\Lambda^{\delta-2}}/{2(\delta-2)}{\Lambda_S^{\delta+2}}$.
The experimental constraints~\cite{vincoli}
give the slightly oblique bound in fig.~\ref{fig:vincoli}.

\item At one loop gravitons affect precision observables and $a_\mu$ in a way that again depends on the cutoff.
The green line shows the values necessary to produce the observed anomaly in $a_\mu$.
The bound parallel to it comes from precision observables.
\end{itemize}
If the cutoff $\Lambda$ is due to quantum gravity,
$\Lambda/M_D$ parameterizes how strongly coupled gravity is:
this explains why virtual graviton effects give the strongest (weakest) bound when 
$\Lambda\circa{>} M_D$ ($\Lambda\circa{<} M_D$).
Strongly coupled gravity is obtained for $\Lambda/M_D\sim (1\div 4)$ if $\delta=3$ and
for $\Lambda/M_D\sim (1\div 2)$ if $\delta=6$.
EWPO bounds have been estimated in a conservative way, assuming a typical $0.1\%$ error.
We see that setting $\Lambda = M_D$ as  assumed in many analyses
is a significant but arbitrary restriction:
$\Lambda$ is a relevant free parameter.
In the most generic case the cutoff could even be not universal,
so that different corrections are cut off by different $\Lambda$.
We repeat that bounds that depend on $\Lambda$ can at best be considered as semi-quantitative.

The presence of a cut-off $\Lambda$ can have an impact on the studies of
graviton emission at future colliders. If $\Lambda$ is smaller than $\sqrt{s}$, real graviton 
signals are suppressed  (but some new physics should show up). On the other hand, if $\Lambda/M_D$
is too big, real graviton signals ($\gamma +$ missing energy) are forbidden by precision 
tests or
subdominant with respect to $\gamma +$ missing energy effects due to dimension six operators like $e\bar{e}\nu\bar{\nu}$~\cite{GRW},
generated by virtual gravitons at one loop with coefficient 
$\sim \pi^2\Lambda^{2\delta+2}/\Lambda_S^{2\delta+4}$. However
there exists a range of $M_D$ and $\Lambda$ (not too small and not too large)  where real
graviton emission is the dominant discovery mode. For instance one can see this by considering the case
of a $e^+e^-$ collider at ${\sqrt s}=1$ TeV \cite{GRW}.

\medskip

{\em Can the apparent excess $a_\mu^{\rm exp} - a_\mu^{\rm SM} = (4.3\pm 1.6) \cdot 10^{-9}$ 
recently measured by~\cite{g-2} be produced by gravitons without conflicting with the EWPO bounds?}
In the SM, electroweak corrections have been clearly seen in the $\epsilon_i$,
but only affect $a_\mu$ at a level comparable to its present experimental error.
The naive (and maybe correct) expectation is that even in the gravitational case
the $\epsilon_i$ are a more significant probe than $a_\mu$.
However, taking into account that we can only perform estimates,
it could not be impossible that the anomaly in $a_\mu$~\cite{g-2} be produced by gravity
without conflicting with the EWPO bounds,
even if the physical cutoff has a `universal' nature (for example if it is related to the size of the brane)
as assumed in fig.~\ref{fig:vincoli}.
If this is the case, 
improved measurements of the $\epsilon_i$ parameters should be able to find a positive signal.

\subsection{Electroweak precision observables}

As discussed in the previous sections, unphysically large corrections cancel out 
when correctly computing physical observables. Previous analyses have studied
certain combinations of the vacuum polarizations of the vector bosons
\begin{equation*}
\Pi^{ij}_{\mu\nu}(k^2) = -i \eta_{\mu\nu}\Pi^{ij}(k^2) + k_\mu k_\nu~{\rm terms},\qquad
i,j = \{W,Z,\gamma\}
\end{equation*}
known as $S,T,U$ parameters~\cite{STU}, often employed to parameterize new physics present only in the
vector boson sector.
However these are not physical observables because gravity does not couple only to vector 
bosons\footnote{Various studies on
different new physics scenarios use the $S,T,U$ approximation outside its domain of applicability.}.
As found in~\cite{EWPOgr1}, 
in the unitary gauge gravitons give corrections to such parameters that unphysically increases with increasing $M_D$.

Since graviton loops are flavour universal (and neglecting the bottom 
quark mass)
gravitational corrections to the various EWPO can be condensed in three parameters
that are usually chosen to be $\epsilon_1,\epsilon_2,\epsilon_3$.
The corrections to the physical EWPO are obtained by combining
in a non immediate  but standard way~\cite{epsilon} various form factors.
Specializing the general expressions to the case of gravity, the $\epsilon$ parameters are given by
\begin{align}
\epsilon_1 =& 2 {\delta g} - \frac{\delta G}{G} - \frac{\delta M_Z^2}{M_Z^2} - \Pi'_{ZZ}(M_Z^2)\\
\epsilon_2 =& 2c^2 {\delta g}-\frac{\delta G}{G}- \frac{\delta M_W^2}{M_W^2} + 
 s^2 \frac{\delta \alpha}{\alpha}-c^2 \Pi'_{ZZ}(M_Z^2)\\
\epsilon_3 =& 2c^2  {\delta g}-c^2 \frac{\delta \alpha}{\alpha}-c^2  \Pi'_{ZZ}(M_Z^2)
\end{align}
where
\begin{itemize}
\item $\delta M_i^2 \equiv -\Pi_{ii}(M_i^2)$ are the correction to the pole mass of
the vector bosons and $\Pi'(k^2)\equiv d\Pi(k^2)/dk^2$.

\item $\delta\alpha=-\Pi_{\gamma\gamma}(0)$ is the correction to the electric charge;

\item $\delta g$ is the common correction to the vector and axial form factors (gravity respects parity)
in the $Z_\mu f\bar{f}$ interactions of an on-shell $Z$ boson 
$$-i\frac{e}{2sc}\bar{f}\gamma_\mu ( g_V - \gamma_5 g_A)(1+\delta g) f$$
excluding the contribution from the $Z$ vacuum polarization.

\item $\delta G$ is the correction to the $\mu\to e\bar{\nu}_e\nu_\mu$ decay amplitude.

\item $s^2\equiv 1-c^2\equiv  [1-\sqrt{1-4\pi\alpha/\sqrt{2}GM_Z^2}]/2 \approx 0.2311$
\end{itemize}
Although it would be straightforward  to perform a complete analysis,
we will only study the gravitational correction to the combination
\begin{equation}
\bar\epsilon\equiv \epsilon_1-\epsilon_2- \frac{s^2}{c^2}\epsilon_3= \frac{\delta M_W^2}{M_W^2} - \frac{\delta M_Z^2}{M_Z^2}
\end{equation}
chosen because it only involves the simplest-to-compute form factors.
Physically, this observable amounts to testing the tree level SM prediction $M_W = c M_Z$ using the value
of the weak angle given by the forward-backward asymmetries in $Z\to \ell^+\ell^-$ decays,
not affected by graviton loop effects.
The experimental value of $\bar\epsilon$ (obtained from a fit of LEP and SLD data) is $\bar\epsilon=(12.5 \pm 1)10^{-3}$
and agrees with the SM prediction (for a light higgs).
The gravitational correction is given in appendix C in terms of Passarino-Veltman functions.
Since the heaviest KK give the dominant effect,
 we can explicitly write the graviton effect
in the limit $m_n\gg M_Z$ as
$$\delta \bar{\epsilon}\approx \sum_n  \frac{s^2 M_Z^2}{\bar{M}_{4}^2(4\pi)^2}\left[
\frac{40+25\delta}{6+3\delta}(\frac{1}{\epsilon}+\ln\frac{\bar\mu^2}{m_n^2})+
\frac{424+546\delta+137\delta^2}{18(2+\delta)^2}\right]$$
where we set $d=4-2\epsilon$.
We can estimate the graviton correction by keeping only the logarithmic term, setting $\bar \mu =\Lambda$
and cutting off the sum at $n< R \Lambda$. This gives
\begin{equation*}
  \delta \bar\epsilon \approx \frac{s^2  M_Z^2}{M_D^2}\bigg(\frac{ \Lambda}{M_D}\bigg)^\delta
   \frac{5(8+5\delta)}{48 \Gamma(2+\delta/2)\pi^{2-\delta/2}}.
\end{equation*}
The result has a strong dependence on $\Lambda$
and the numerical coefficient is specific of the form of the cutoff that we have chosen to employ.
We explicitly see that spurious IR divergences do not affect this physical observable.

Notice that $\bar{\epsilon}$ is suppressed by a power of $s^2$ but it is not affected by
theoretical uncertainties in $\delta\alpha$.
Therefore precision searches for $M_D$ could be improved by a factor $\sim 3$ if,
by producing $\sim 10^9$ $Z$ bosons at an $e\bar{e}$ linear collider, the
errors on $M_W$ and on the effective weak angle extracted from the leptonic asymmetries
could be reduced by a factor $\sim 10$.

\subsection{Anomalous magnetic moment of the $\mu$}

Since the $\mu$ anomalous magnetic moment is zero at tree level, the reparametrization
gauge dependence of the unit of mass does not affect the
one loop gravitational correction to $a_\mu$.
Only few Feynman diagrams contribute. As noticed in~\cite{antico}, 
the $1/\epsilon$ poles cancel out
when computing the loop integrals using dimensional regularization around $d=4$.
At leading order in $m_\mu$  we find, again using a sharp cut off for the sum over KK modes
\begin{equation*}
\delta a_\mu = \frac{m_\mu^2}{M_D^2}   \bigg(\frac{ \Lambda}{M_D}\bigg)^\delta\frac{ 34 + 11\delta}
{96 \Gamma(2 + {\delta }/{2})\pi^{2-\delta/2}}.
\end{equation*}
Apparently this result agrees with the one found by~\cite{grasser}\footnote{\cite{grasser} separately computes
the gauge-dependent `graviton' and `radion' contributions in the unitary gauge.
We find a different result in both cases
(the radion coupling used in~\cite{grasser} is valid only on shell),
but this discrepancy luckily cancels out when summing the two contributions.}.
Again the result {\em strongly} depends on the value of the cutoff $\Lambda$.
We cannot claim that it has the same sign as
the apparent excess  recently measured by~\cite{g-2}:
we have employed dimensional regularization for loop integrals but other regularizations
(e.g. dimensional reduction, Pauli-Villars,\ldots) would give a different result.
Unlike $\bar\epsilon$, $\delta a_\mu$ is a sum of contributions from graphs with different graviton interactions.
One can obtain any sign for $\delta a_\mu$ e.g.\ by cutting off
 $\bar{\mu}\mu h$ and $\gamma\gamma h$
vertices with different form factors: $\delta a_\mu$ is finite but dominated
by loop momenta around the cutoff.
In particular one gets, at leading order in $m_\mu$
\begin{equation*}
\delta a_\mu=0
\end{equation*}
if the cutoff acts in the same way on both type of contributions.
This is for example the case of a Pauli-Villars cutoff on the graviton.
By working in the De Donder gauge (where  the only dependence on the graviton mass
comes from the $1/(k^2-m_n^2)$ factor in the graviton propagator) and knowing that $a_\mu$ is dimensionless 
and finite, it is not difficult to realize that it is zero.

\section{Conclusions}

We  discussed various subtleties that arise when computing quantum gravity
1-loop effects in models with
large extra dimensions and matter confined to a brane. Our computations are based 
on an effective field theory (EFT) description of quantum gravity and of the brane.
A sensible result is obtained after correctly identifying physical
observables and after taking  brane fluctuations into account. 
Graviton tadpoles are relevant for branes living in non-homogeneous spaces (like orbifolds).
For branes living at orbifold
fixed points consistency is met, as  expected, even in the absence of brane fluctuations.
In particular we explain in a geometric way why
the units of length in `longitudinal' and `transverse' directions depend
on the reparametrization gauge fixing procedure.

We regard these results as theoretically interesting, although the truly calculable
effects in the EFT approach have a limited phenomenological relevance.
The most relevant effects come from the region of large virtual momenta
where the EFT description breaks down. 
This is why  in the second part of the paper we have abandoned the strict EFT approach
and modeled these UV effects by introducing a hard momentum cut-off $\Lambda$. This is the best that 
can be done, without having a fundamental theory
that allows real computations. We stress however
that our previous understanding of how to get gauge independent results
is still important in this phenomenological approach.
As an application we have studied virtual graviton corrections to
precision observables and to the muon anomalous magnetic moment,
focusing on models with large extra dimensions.
Even at tree level,  virtual graviton effects are divergent and must be regulated.
%so that we
%estimate them  by introducing a hard momentum cut-off $\Lambda$.
Virtual graviton effects in collider phenomenology have been so far studied assuming a
particular value of $\Lambda$.
However $\Lambda$ is an important free parameter that --- at least at
an qualitative level --- controls how strongly coupled gravity is.
Depending on the value of $\Lambda$, one loop effects can give the dominant
bound on low scale quantum gravity.

\acknowledgments
We would like to thank Ignatios Antoniadis, Gian Giudice and Angel Uranga for
interesting discussions.
R.C.\ is indebted with G.Policastro, R.Sturani and especially with A.Gambassi, M.Gubinelli 
and A.Montanari for many useful discussions. This work is partially supported
by the EC under TMR contract HPRN-CT-2000-00148.

\appendix

\section{Graviton propagators in the unitary gauge}\label{unitary}

We derive here the propagators for the physical fields in the unitary gauge.
Expanding the metric $g_{MN}=\eta_{MN}+\kappa h_{MN}$ around the flat space solution we obtain the
Lagrangian in eq.\eq{linear}.
We recall that we decompose the $D$ dimensional graviton $h_{MN}$ as $h_{\mu\nu}$, $h_{\mu i}$ and $h_{ij}$
(where $\mu,\nu$ are $4$-dimensional indices and $i,j$ span the extra $\delta$ dimensions).
In this appendix we fix for simplicity $d=4$.
Due to $D$-dimensional reparametrization invariance not all the components of these fields correspond to propagating degrees of freedom.
The physical fields are the ones contained in the Riemann tensor $2R_{RMSN}= h_{(RN,MS)}-h_{(MN,RS)}$
(as in the electromagnetic case the physical fields are 
contained in the field strength tensor)
\begin{align}
G_{\mu\nu} =& -2 {\partial}_{\alphad}{\partial}_{\betad} R_{\alphad\mu\betad\nu}=
h_{\mu\nu}-{\partial}_{i}\partial_{(\mu} h_{\nu)i}+ \partial_\mu\partial_\nu {\partial}_{\alphad}{\partial}_{\betad} h_{\alphad\betad}\\
V_{\mu\alphad} =&+ 2{\partial}_\betad {\partial}_\nud R_{\betad\mu\alphad\nud}= h_{\mu\alphad} -
{\partial}_{\alphad}{\partial}_{\nud} h_{\mu\nud}-
\partial_\mu {\partial}_\betad h_{\alphad\betad} + \partial_\mu {\partial}_\alphad {\partial}_\betad {\partial}_\nud
h_{\betad\nud}\\ S_{\alphad\betad} =& -2{\partial}_\mud {\partial}_\nud R_{\alphad\mud\betad\nud} =h_{\alphad\betad}-{\partial}_{n}
{\partial}_{(\alphad}h_{\betad) n}+
{\partial}_\alphad {\partial}_\betad {\partial}_\mud {\partial}_\nud h_{\mud\nud}
\end{align}
For simplicity, the above equations are written assuming units such that $\partial_i\partial_i=1$.
These expressions can be written in a compact form by defining
$Q_\mu\equiv \hat\partial_i h_{\mu i}$, $P\equiv \hat\partial_i \hat\partial_j h_{ij}$ and $P_i=\hat\partial_j h_{ij}-\hat\partial_i P$,
where $\hat{\partial}_i\equiv \partial_i/\sqrt{\partial_j \partial_j}$.
These considerations suggest to rewrite the Lagrangian in terms of a new set of fields \cite{GRW}
$G_{\mu\nu}$, $V_{\mu i}$, $S_{ij}$, $H$,
$Q_\mu$, $P_i$,
$P$ related to $h_{MN}$ by
\begin{align} \label{eq:h=G+H}
h_{\mu\nu} =& G_{\mu\nu}-\frac{c}{4-1}(\eta_{\mu\nu}+\frac{\partial_\mu \partial_\nu}{\partial_i^2}) H+
\partial_\mu Q_\nu + \partial_\nu Q_\mu- \partial_\mu \partial_\nu P \\
h_{ij} =& S_{ij} + \frac{c}{\delta-1}(\eta_{ij}-\hat\partial_i \hat\partial_j) H + 
\hat\partial_i P_j + \hat\partial_j P_i + \hat\partial_i \hat\partial_j P\\
h_{\mu i}  =& V_{\mu i} + \partial_\mu P_i + \hat\partial_i Q_\mu
\end{align}
and subject to the constraints
$$\partial_i V_{\mu i}= \partial_i S_{ij} = \partial_i P_i =0,\qquad
S_{ii}=0$$
We have introduced the field $H$ in order to make $S_{ij}$ traceless. By choosing
 $c^2=3(\delta-1)/(\delta+2)$, $H$ is canonically normalized.
The $D$-dimensional Lagrangian becomes
\begin{equation} \label{eq:Ldiagonal}
\begin{split}
{\cal L} =&-\frac{1}{2} H(\QQ+\partial_k^2) H - \frac{1}{2} S^{ij}(\QQ+\partial_k^2)S_{ij}-
 V^{\mu i}[(\QQ+\partial_k^2)\eta_{\mu\nu}-\partial_\mu\partial_\nu]V^{\nu}_i+\\
 &\frac{1}{2} G^\mu_\mu(\QQ+\partial_k^2)G^\nu_\nu-\frac{1}{4} G^{\mu\nu}(\QQ+\partial_k^2) G_{\mu\nu}+
 G^{\mu\rho}\partial_\rho \partial_\sigma G_\mu^\sigma- G_\rho^\rho\partial_\mu\partial_\nu
 G^{\mu\nu}.
\end{split}
\end{equation}
As expected, it does not depend on $Q_\mu$, $P_i$ and $P$ and there is no mixing between the 
fields $G_{\mu\nu}$, $H$, $S_{ij}$, $V_{\mu i}$.
It is now trivial to perform a mode expansion: the extra dimensional Laplacian $\partial_k^2$ becomes 
a mass term. The propagators can be obtained by inverting the kinetic terms in eq.~(\ref{eq:Ldiagonal}).
It is useful to show explicitly how the `graviton' $G_{\mu\nu}$ and the `scalar' $H$ combine to give
a unitary-gauge propagator equal to the de-Donder propagator,
up to longitudinal terms.
For example, the $h_{\mu\nu}h_{\mu'\nu'}$ propagator is
\begin{equation}\label{eq:h-prop}
\begin{split}
\Tcontr{0.4em}{2.5ex}{1ex}{1.6em}h_{\mu\nu}^{(n)}h_{\mu'\nu'}^{(n')} = \, &
 \Tcontr{0.4em}{2.5ex}{1ex}{1.8em}G_{\mu\nu}^{(n)}G_{\mu'\nu'}^{(n')} + 
 \frac{\delta-1}{3(\delta+2)}t_{\mu\nu}t_{\mu'\nu'}\Tcontr{0.4em}{2.5ex}{1ex}{1.8em}H^{(n)}H^{(n')}=\\
&
\frac{i\delta_{n,-n'}}{2(k^2-\Mn^2)}
 \Big(t_{\mu\mu'}t_{\nu\nu'}+t_{\mu\nu'}t_{\nu\mu'}-\frac{2}{2+\delta}t_{\mu\nu}t_{\mu'\nu'}\Big)
\end{split}
\end{equation}
In particular we see that we cannot omit the `scalar' contributions,
if we want to obtain a gauge invariant result.
It would be easy to include a small mass term for the $H^{(n)}$ fields,
eventually generated by the unknown mechanism that stabilizes the size of the extra dimensions.

\section{Graviton vertices}\label{vertici}

We define $g_{\mu\nu}\equiv \eta_{\mu\nu}+\kappa \, h_{\mu\nu}$, $g\equiv |\det g_{\mu\nu}|$ and
give explicit expressions for the expansion up to second order in the graviton field $h_{\mu\nu}$ of 
\begin{align}
\sqrt{g} =& 1 + \kappa A^{\alpha\beta} h_{\alpha\beta} + 
\kappa^2 A^{\prime \alpha\beta\gamma\delta} h_{\alpha\beta}h_{\gamma\delta}+\cdots\\
\sqrt{g}g^{\mu\nu} =& \eta^{\mu\nu} + \kappa B^{\mu\nu\alpha\beta} h_{\alpha\beta} + 
\kappa^2 B^{\prime \mu\nu\alpha\beta\gamma\delta}
h_{\alpha\beta}h_{\gamma\delta}+\cdots\\
\sqrt{g}g^{\mu\rho} g^{\nu\sigma} =& \eta^{\mu\rho} \eta^{\nu\sigma} + \kappa C^{\mu\nu\rho\sigma\alpha\beta} h_{\alpha\beta} + 
\kappa^2 C^{\prime \mu\nu\rho\sigma\alpha\beta\gamma\delta}
h_{\alpha\beta}h_{\gamma\delta}+\cdots
\end{align}
In the vierbein formalism the spin connection is given by
\begin{equation} \label{eq:e2omega}
{\omega_a}^{cd} \, = \, e^\mu_a \, {\omega_\mu}^{cd} \, = \, 
 e^\mu_a \left(e^{\nu c} \, \partial_{[\mu} e^d_{\nu]} \, - \, e^{\nu d} \, \partial_{[\mu} 
 e^c_{\nu]} \right) \, - \, e^{\rho c} \, e^{\sigma d} \, \partial_{[\rho} e^m_{\sigma]} \, \eta_{ma} \; ,
\end{equation}
We expand around the flat background $\delta^a_\mu$, $e^a_\mu = \delta^a_\mu + \, \kappa \,  b^a_\mu$. 
As discussed in  section~\ref{fermions},
the gauge choice $b_{[\mu\,\nu]}=0$ allows to express $b_{\mu\nu}$ 
in terms of $h_{\mu\nu}$
\begin{equation} \label{eq:B2h}
b_{\mu \nu} \, = \, \frac{1}{2} \, h_{\mu \nu} \, - \,  \frac{\kappa}{8} \, h^\alpha_\mu \, h_{\alpha \nu}
\, + \, O(\kappa^2)
\end{equation}
Using (\ref{eq:e2omega}),~(\ref{eq:B2h}) and 
\begin{equation}
e^\mu_a \, = \, \delta^\mu_a \, -\, \kappa \, b^\mu_a  \, + \, \kappa^2 \,   
 b^\mu_\alpha b^\alpha_a \, + \, O(\kappa^3) 
\end{equation}
one can find the gravitational couplings for fermions.

With these expressions it is straightforward to find the graviton vertices arising from Lagrangians like
\begin{equation*}
\begin{split}
{\cal L} =
 \sqrt{g} \bigg[ &g^{\mu\nu}\frac{(\partial_\mu \phi)(\partial_\nu \phi)}{2}-\frac{m^2_\phi}{2} \phi^2-
 \frac{1}{4} g^{\mu\rho}g^{\nu\sigma}F_{\mu\nu}F_{\rho\sigma}- 
 \frac{m_A^2}{2} g^{\mu\nu}A_\mu A_\nu + \\  
 & \frac{i}{2} \left( \overline{\psi} \, e^\mu_a \, \gamma^a \, D_\mu \psi \, - \,  
 D^\dagger_\mu  \overline{\psi}  \, e^\mu_a \, \gamma^a  \, \psi \right)\bigg]
\end{split}
\end{equation*}
where $F_{\mu\nu} \equiv \partial_\mu A_\nu - \partial_\nu A_\mu$ and 
$D_\mu \psi \, = \, \partial_\mu \psi \, + \, \frac{1}{2} {\omega_\mu}^{ab} \, \gamma_{ab} \, \psi$,
$\gamma^{ab} = \frac{1}{4} [\gamma^a,\gamma^b]$.
The expansion in powers of $h$ is easily obtained from
\begin{align}
g^{\mu\nu} =& \eta^{\mu\rho} [\delta^{\nu}_{\rho}- \kappa h^{\nu}_{\rho} + 
 \kappa^2 (h  h)^{\nu}_{\rho}-\kappa^3 (hhh)^{\nu}_{\rho}+\cdots]\\[0.1cm]
\sqrt{g}  =& 1+\kappa\frac{\Tr h}{2}+
\kappa^2\bigg[ \frac{\Tr^2 h}{8}-\frac{\Tr h^2}{4}\bigg]+
\kappa^3\bigg[ \frac{\Tr^3 h}{48}-\frac{\Tr h \Tr h^2}{8}+\frac{\Tr h^3}{6}\bigg]+\cdots
\end{align}
Therefore
\begin{align}
A^{\alpha\beta} =& \frac{1}{2} \eta^{\alpha\beta}\\
 B^{\mu\nu\alpha\beta} =& 
4 A^{\prime \mu\nu\alpha\beta} =\frac{1}{2} 
  (\eta^{\mu\nu}\eta^{\alpha\beta}-\eta^{\mu\alpha}\eta^{\nu\beta}-
   \eta^{\mu\beta}\eta^{\nu\alpha}) \\
B^{\prime \mu\nu\alpha\beta\gamma\delta} =&  \frac{1}{4} \eta^{\mu\nu} B^{\alpha\beta\gamma\delta} -
  \frac{1}{2} \Big( \eta^{\mu\alpha} B^{\gamma\delta\nu\beta}+
  \eta^{\nu\alpha} B^{\gamma\delta\mu\beta}\Big) \\
C^{\mu\nu\rho\sigma\alpha\beta} =&   \Big[
 \frac{1}{2} \eta^{\mu\rho}\eta^{\nu\sigma}\eta^{\alpha\beta}
    -\eta^{\mu\alpha}\eta^{\beta\rho}\eta^{\nu\sigma}
    -\eta^{\nu\alpha}\eta^{\beta\sigma}\eta^{\mu\rho}  \Big] \\
C^{\prime \mu\nu\rho\sigma\alpha\beta\gamma\delta}=&
 \Big[ \frac{1}{4} B^{\alpha\beta\gamma\delta}
  \eta^{\mu\rho}\eta^{\nu\sigma}
 +2 \eta^{\gamma\mu}\eta^{\alpha\rho}\eta^{\delta\beta}\eta^{\nu\sigma}
 -\eta^{\gamma\delta}\eta^{\alpha\mu}\eta^{\beta\rho}\eta^{\nu\sigma}
 +  \eta^{\alpha\mu}\eta^{\beta\rho}\eta^{\gamma\nu}\eta^{\delta\sigma} \Big]
\end{align}
%
% \begin{align}
% A^{\alpha\beta} =& \frac{1}{2} \eta^{\alpha\beta}\\
%  B^{\mu\nu\alpha\beta} =& 
% 4 A^{\prime \mu\nu\alpha\beta} =\frac{1}{2} 
%   (\eta^{\mu\nu}\eta^{\rho\sigma}-\eta^{\mu\rho}\eta^{\nu\sigma}-
%    \eta^{\mu\sigma}\eta^{\nu\rho}) \\
% B^{\prime \mu\nu\alpha\beta\gamma\delta} =&  \frac{1}{2} \eta^{\mu\nu} B^{\rho\sigma\alpha\beta} -
%   \Big( \eta^{\mu\rho} B^{\alpha\beta\nu\sigma}+
%   \eta^{\nu\rho} B^{\alpha\beta\mu\sigma}\Big) \\
% C^{\mu\nu\rho\sigma\alpha\beta} =&   \Big[
%  \frac{1}{2} \eta^{\mu\rho}\eta^{\nu\sigma}\eta^{\alpha\beta}
%     -\eta^{\mu\alpha}\eta^{\beta\rho}\eta^{\nu\sigma}
%     -\eta^{\nu\alpha}\eta^{\beta\sigma}\eta^{\mu\rho}  \Big] \\
% C^{\prime \mu\nu\rho\sigma\alpha\beta\gamma\delta}=&
%  \Big[ \frac{1}{4} B^{\alpha\beta\gamma\delta}
%   \eta^{\mu\rho}\eta^{\nu\sigma}
%  +2 \eta^{\gamma\mu}\eta^{\alpha\rho}\eta^{\delta\beta}\eta^{\nu\sigma}
%  -\eta^{\gamma\delta}\eta^{\alpha\mu}\eta^{\beta\rho}\eta^{\nu\sigma}
%  +  \eta^{\alpha\mu}\eta^{\beta\rho}\eta^{\gamma\nu}\eta^{\delta\sigma} \Big]
% \end{align}
%
These expressions are valid in any number of dimensions.
Brane fluctuations can be incorporated in $h_{\mu\nu}$, as discussed in eq.~\eq{heff}.

To compute the corrections to the graviton propagator and to the graviton vertex
it is necessary to have the 3 and 4 graviton interactions.
They can be easily derived by expanding the Einstein-Hilbert Lagrangian
in powers of the graviton field using e.g. Mathematica~\cite{Mathematica}.
For this reason we do not write explicitlt the long expressions for such vertices.

\section{Results}\label{risultati}

In this appendix we collect the explicit results for the corrections to the propagator
of a spin 0,1 particle confined on a brane with dimension $d$ living in $\mathbb{R}^d \times T^\delta$.
As discussed in sec.~3, generically the correction to a
physical quantity ${\cal O}$ with  canonical dimension $d_{ {\cal O}}$ has the form
\begin{equation}
\delta {\cal O}/{\cal O} = d_{ {\cal O}} \, G(M_DR) + \Delta ({\cal O},R,M_D,\mu)
\end{equation}
where the gauge dependence is encoded in the function $G$.
The splitting in a `gauge-dependent' and `gauge-independent' part is ambiguous 
unless a reference gauge is chosen in which by definition one sets $G_{\text{ref}}=0$.
We choose in the de Donder gauge $G_{\text{de Donder}}=0$.
All the results for physical quantities are computed in this gauge.
The results are expressed in terms of Passarino-Veltman functions $A_0$, $B_{0,1}$, 
defined in appendix~\ref{integrals}.

For the pole mass correction for a scalar ($s=0$) and a massive vector ($s=1$) on the brane we find
\begin{equation}
G_{\rm unitary} = -\sum_{n} A_0(\Mn^2) \,
          \frac{\big[ d^2 + (\delta-3)\delta +d(2\delta-1)\big]}{16\pi^2 \bar M_d^{d-2}d(d+\delta-2)} 
\end{equation}
\begin{align}
\begin{split}
\Delta_0 &=  \frac{1}{64 \pi^2 \bar M_d^{d-2} d (d+\delta-2)} \sum_{n} \Big\{ 
          2d(2-d)(\delta-4) A_0(m_0^2) + f_1(d,\delta) A_0(\Mn^2)+ \\
        & 8d \big[ \Mn^2 (d-2) - 2 m_0^2  (d+\delta-3)\big] 
          B_0(m_0^2,\Mn^2,m_0^2) + 2d\delta (d-2)  \Mn^2 B_1(m_0^2,\Mn^2,m_0^2) \Big\} 
\end{split} \\
\begin{split} 
\Delta_1 &= \frac{1}{64 \pi^2 \bar M_d^{d-2} d (d-1) (d+\delta-2)} \sum_{n} \Big\{ 
           f_2(d,\delta) A_0(m_1^2) + (d-1) f_1(d,\delta) A_0(\Mn^2) - \\
         &8d \big[ 2 m_1^2 (d-1)(d+\delta-3) + \Mn^2 (d-2)(d+2\delta+1) \big] B_0(m_1^2,\Mn^2,m_1^2) + \\
         &2 d(d-2) \big[2d^2+3d(\delta-2)-7\delta\big] \Mn^2  B_1(m_1^2,\Mn^2,m_1^2) \Big\}
\end{split}
\end{align}
with
\begin{equation}
\begin{split}
f_1(d,\delta) &= 4\big[2d^3-d(\delta-6)+2d^2 (\delta-3)+ 2\delta (\delta-2)\big] \\
f_2(d,\delta) &= 2d (2-d) \big[4+2d^2+\delta+d(3\delta-2)\big]
\end{split}
\end{equation}
We computed also the graviton correction to the photon propagator,
verifying that it is transverse if one uses the simple gauge fixing of eq.~(\ref{eq:simpleGF}).
If instead the gauge fixing function contains the graviton field, in general transversality will be
lost, due to a modification of the related Ward identity (of course this does not mean that the 
photon acquires a mass). 
As the simple QED case, Ward identities imply that the photon self energy at zero momentum gives
the correction to the electric charge. We find
\begin{equation}
\Delta_{e} = -\sum_{n} A_0(\Mn^2) \, 
 \frac{(d-4)\big[d^3+ d^2 (\delta-5)+ d (8-3\delta) 
  +2\delta (\delta+2)\big]}{32 \pi^2 
 \bar M_d^{d-2} d (d+\delta-2)}  
\end{equation}
For comparison with the existing literature, we also write the expression of the scalar self-energy 
at zero momentum $\Sigma(0)$. As discussed in the text, for this unphysical quantity the gauge dependent
part is non-universal. In the de Donder and in the unitary gauge we find
\begin{equation}
\begin{split}
\Sigma(0)_{\text{de Donder}} &= 
 \frac{m_0^2}{32 \pi^2 \bar M_d^{d-2} (d+\delta-2)} \sum_{n} \, \frac{1}{(m_0^2-\Mn^2)} \\
    &  \Big\{ \big[ m_0^2 g_1(d,\delta)+\Mn^2 g_2(d,\delta) \big] A_0(\Mn^2) + 
 2d (\delta-2) m_0^2 A_0(m_0^2)\Big\}
\end{split}
\end{equation}
\begin{equation}
\begin{split} 
\Sigma(0)_{\text{unitary}} &=
 \frac{m_0^2}{32 \pi^2  \bar M_d^{d-2}(d+\delta-2)} 
 \times  \sum_{n} \, \frac{1}{\Mn^4 (m_0^2-\Mn^2)} \\
     &  \Big\{ 2m_0^2 \big[ \Mn^4 d(\delta-2) -2\Mn^2 m_0^2 (\delta-2) +
 m_0^4 (d+\delta-3)\big] A_0(m_0^2) + \\
     &(d-1) \Mn^4 \big[ m_0^2 \big( d^2+d(\delta-2)-2\big) -
 \Mn^2 (d-2)(d+\delta)\big] A_0(\Mn^2) \Big\}
\end{split}
\end{equation}
with
\begin{equation}
\begin{split}
g_1(d,\delta) &= d^2(d-1) -4d +\delta(4+d)(d-1)+2\delta^2 \\
g_2(d,\delta) &= 4\delta - (d+\delta) \big[2\delta +d (d-1) \big]
\end{split}
\end{equation}
From these expressions it is clear that $1/\Mn^4$ terms found in \cite{EWPOgr1,EWPOgr2,Bij} 
are an artifact of the unitary gauge and have no physical meaning.

\section{Regularized sums and integrals}\label{integrals}

The results of our 1-loop computations  can be 
expressed as the sum over KK modes of basic Passarino-Veltman functions~\cite{Passarino}.
Generically these expressions are divergent and need to be regulated. After doing that
one can extract the calculable finite parts that are determined by the EFT \cite{Dono}. 
These are terms that 
either depend on the radius $R$ or depend non-analytically on the kinematic variables.
In this appendix we focus for illustration on these calculable terms and disregard
the uncalculable UV saturated contribution, which were the subject of our phenomenological 
discussion.

The main point is to regularize the integral and the series
consistently; we choose for this the dimensional technique,
extending the physical dimension of the extra space and of the brane $\delta$, $d$,
to generic values
\begin{equation*}
\bar\delta = \delta -\epsilon; \qquad \bar d = d-\epsilon
\end{equation*}
rescaling the Planck mass in the Lagrangian as
$M_D^{d+\delta-2}\to M_D^{d+\delta-2}\mu^{\bar d+\bar\delta-d-\delta}$ and consequently
\begin{equation*}
\frac{1}{\bar M_d^{d-2}} \equiv \frac{1}{M_D^{D-2} R^\delta} \to \frac{\mu^{2\epsilon}}{M_D^{D-2} R^{\bar\delta}}
\end{equation*}
and taking the limit $\epsilon\to 0$ at the end.
Defining the Passarino-Veltman functions $A_0$, $B_{0,1}$
\begin{equation}
\begin{split}
A_0(m^2) &=  -i(4\pi)^2  \int \frac{d^{ d}q}{(2\pi)^{ d}} \frac{1}{q^2-m^2}  \\
B_0(p^2,M^2,m^2) &= -i(4\pi)^2  \int \frac{d^{ d}q}{(2\pi)^{ d}} \frac{1}{(q^2-M^2)[(q+p)^2-m^2]} \\
B_1(p^2,M^2,m^2) &= \frac{-i(4\pi)^2}{p^2} 
 \int \frac{d^{ d}q}{(2\pi)^{ d}} \frac{p\cdot q}{(q^2-M^2)[(q+p)^2-m^2]} 
\end{split}
\end{equation}
the following expressions are sufficient to compute the gravitational corrections in
appendix~\ref{risultati}
\begin{equation} \label{eq:PVint}
\sum_{n\in \mathbb{Z}^{\delta}} A_0(\Mn^2) \qquad 
 \sum_{n\in \mathbb{Z}^{\delta}} \Mn^{2\alpha} B_{0,1}(m^2,\Mn^2,m^2) \qquad \qquad \alpha=0,1
\end{equation}
together with the series
\begin{equation} \label{eq:series}
{\cal I}_\alpha = \sum_{n\in \mathbb{Z}^{\delta}} \frac{1}{n^{2\alpha}}
\end{equation}
To illustrate the technique we compute explicitly $\sum_n B_0 (m^2,\Mn^2,m^2)$.
First of all we introduce a Feynman parameter $x$, rescale the integration variable
$q\to q/R$ and isolate the zero point in the series
\begin{equation} \label{eq:starting}
\begin{split}
\sum_{n\in \mathbb{Z}^{\bar\delta}} &B_0(m^2,\Mn^2,m^2) = \\
  &-\frac{4i}{(2\pi)^{\bar d-2}} \int_0^1 dx \Big[ \sum_{n\in\mathbb{Z}^{\bar\delta}-\{0\}} \int d^{\bar d}q
  \frac{(\sqrt{x}/R)^{\bar d-4}}{[q^2-n^2-a^2(x)]^2} + \int d^{\bar d}q \frac{1}{[q^2-(1-x)^2 m^2]^2} \Big]
\end{split}
\end{equation}  
where $a^2(x)=R^2 m^2 (1-x)^2/x$.
Then we Wick-rotate and evaluate the first term  using the Schwinger's proper time method
\begin{equation} \label{eq:intermedio}
\begin{split}
\sum_{n\in\mathbb{Z}^{\bar\delta}-\{0\}} \int d^{\bar d}q \frac{1}{[q^2+n^2+a^2(x)]^2} 
 =& \sum_{n\in\mathbb{Z}^{\bar\delta}-\{0\}} \int^{\infty}_0 dt \int d^{\bar d}q \frac{t}{\Gamma(2)}
    e^{-t(q^2+n^2+a^2)} \\
 =&  \pi^2 \int_0^\infty dy \big[{\cal B}^{\bar\delta}(y)-1\big] e^{-y\pi a^2} y^{1-{\bar d}/2}
\end{split}
\end{equation}
where we have performed the gaussian integral over $q$ and introduced the special function
\begin{equation}
{\cal B}(s) \equiv \sum_{n=-\infty}^{\infty} e^{-\pi n^2 s}
\end{equation}
The integral in eq.~(\ref{eq:intermedio}) converges at $y\to \infty$ thanks to the exponential
behavior of the ${\cal B}$ function, but it diverges at $y\to 0$.  To extract the 
singularity it is useful the property
\begin{equation} \label{eq:Bfun}
{\cal B}(s) = s^{-1/2} {\cal B}(\frac{1}{s})
\end{equation}
which is easily derived from the Poisson formula.
Using eq.~(\ref{eq:Bfun}) we can split the integration interval and change variable $y\to 1/y$ 
in the first integral
\begin{equation}
\begin{split}
 {\cal I} \equiv & \Big( \int^1_0  +\int^{\infty}_1 \Big)
          dy \; y^{1-{\bar d}/2} e^{-y\pi a^2} \big[{\cal B}^{\bar\delta}(y)-1\big] \\ 
 =& \int^{\infty}_1 dy \; \big[{\cal B}^{\bar\delta}(y)-1\big] 
    \big( y^{1-{\bar d}/2} e^{-y\pi a^2} +y^{(\bar d+\bar\delta-3)/2} e^{-\pi a^2/y} \big) + \\
 \; & \int^{\infty}_1 dy \; e^{-\pi a^2/y} y^{\bar d/2-3} \big(y^{\bar\delta/2}-1 \big) 
\end{split}
\end{equation}  
Again, the first integral is convergent, while the second term must be (dimensionally) regularized.
Because $a(x)\ge 0$ in $x\in[0,1]$ and noting that for $\beta\geq 0$
\begin{equation}
\int^{\infty}_1 dy \; y^\alpha e^{-\beta/y} = \beta^{\alpha+1} \Gamma(-\alpha-1) -
 \int^{\infty}_1 dy \; y^{-(\alpha+2)}  e^{-\beta y}
\end{equation}
we can isolate the divergent piece in the $\Gamma$ function through an analytical continuation
in the physical region $\bar d+\bar\delta\geq 0$
\begin{equation}
\begin{split}
{\cal I}  =& \int^{\infty}_1 dy \; \big[{\cal B}^{\bar\delta}(y)-1\big] 
   \big( y^{1-\bar d/2} e^{-y\pi a(x)^2} +y^{(\bar d+\bar\delta)/2-3} e^{-\pi a(x)^2/y} \big) + \\
 & \int^{\infty}_1 dy \; y^{1-\bar d/2} e^{-\pi y a(x)^2} \big( 1-y^{-\bar\delta/2 }\big) +
   \left[\pi R^2 m^2 \frac{(1-x)^2}{x}\right]^{(\bar d+\bar\delta)/2-2} \Gamma(2-\frac{\bar d+\bar\delta}{2}) - \\ 
 & \left[\pi R^2 m^2 \frac{(1-x)^2}{x}\right]^{\bar d/2-2} \Gamma(2-\frac{\bar d}{2})  
\end{split}
\end{equation}
It's not difficult to verify that the last (divergent) term is exactly canceled by the 
the zero mode contribution of the series in eq.~(\ref{eq:starting}).
Putting together the remaining terms we obtain
\begin{equation} \label{eq:Bgen}
\begin{split}
\sum_{n\in \mathbb{Z}^{\bar \delta}} B_0 (m^2,&\Mn^2,m^2)
 = \frac{1}{(2\pi R)^{\bar d-4}}  \int^1_0 dx \; x^{\bar d/2-2} \\ &\Big\{ f_1(mR,x,\bar d,\bar\delta) + 
\left[\pi R^2 m^2 \frac{(1-x)^2}{x}\right]^{(\bar d+\bar\delta)/2-2} \Gamma(2-\frac{\bar d+\bar\delta}{2})\Big\}
\end{split}
\end{equation} 
where we have defined
\begin{equation}
\begin{split}
f_1(mR,x,d,\delta) = \int_1^\infty dy \; \Big\{ &[{\cal B}^\delta(y)-1 ] 
   \big(y^{1-d/2} e^{-y\pi a^2(x)} + y^{(d+\delta)/2-3} e^{-\pi a^2(x)/y}\big) + \\
  &e^{-y\pi a^2(x)} y^{1-d/2} \big(1-y^{-\delta/2}\big) \Big\}
\end{split}
\end{equation}
The $\Gamma$ function in eq.~(\ref{eq:Bgen}) has poles for negative integer arguments and before
taking the limit $\epsilon\to 0$ we must distinguish the two cases of even and odd $(d+\delta)$.
If $(d+\delta)$ is {\it even}, a logarithmic term appears
\begin{equation}
\begin{split}
F(\bar d,\bar\delta)\frac{\mu^{2\epsilon}}{R^{\bar\delta}} 
 &\sum_{n\in \mathbb{Z}^{\bar\delta}} B_0 (m^2,\Mn^2,m^2) =
  \frac{F(d,\delta)}{(2\pi)^{d-4}} \int dx \, x^{d/2-2} \\ 
  \Big\{ &\frac{1}{R^{d+\delta-4}} \, f_1(mR,x,d,\delta) + 
  \frac{m^{d+\delta-4}}{\Gamma\left((d+\delta)/2-1\right)} 
  \left[ -\frac{\pi(1-x)^2}{x} \right]^{(d+\delta)/2-2}  \\
  &\Big( \frac{1}{\epsilon} +\log \frac{\mu^2}{m^2} +\log\frac{2\sqrt{x}}{(1-x)^2} -
  \gamma_E+\frac{1}{F(d,\delta)}\frac{dF}{d\epsilon}\Big|_{\epsilon=0} \Big) \Big\}
\end{split}
\end{equation}
where $F(d,\delta)$ is a generic function of $d,\delta$ which multiplies the integral in the physical 
amplitudes
and the factor $1/R^{\bar\delta}$ comes from the graviton wave function normalization.
By subtracting just the pole $1/\epsilon$ we get the loop correction in the MS scheme.
Notice that the finite part contains a scheme independent $\log m$ term.
When $(d+\delta)$ is {\it odd} we find instead a finite result
\begin{equation}
\begin{split}
F(\bar d,\bar\delta) \frac{\mu^{2\epsilon}}{R^{\bar\delta}} 
 &\sum_{n\in \mathbb{Z}^{\bar\delta}} B_0 (m^2,\Mn^2,m^2) =
  \frac{F(d,\delta)}{(2\pi)^{d-4}} \int dx \, x^{d/2-2} \\ 
  \Big\{ &\frac{1}{R^{d+\delta-4}} \, f_1(mR,x,d,\delta) + 
  m^{d+\delta-4} \, \Gamma\left(2-\frac{d+\delta}{2} \right) 
  \left[ \frac{\pi(1-x)^2}{x} \right]^{(d+\delta)/2-2}  \Big\}
\end{split}
\end{equation}
Although there is no logarithm, the term $m^{d +\delta-4}$ represents a scheme independent
finite effects as it depends non analytically on the Lagrangian parameter $m^2$.
The same technique can be used to compute the finite part of the other integrals in eq.~(\ref{eq:PVint})
and the series in eq.~(\ref{eq:series}); here we collect only the final results omitting the derivation 
(for $d>2$, $\alpha<\delta/2$)
\begin{gather}
\sum_{n\in \mathbb{Z}^{\bar\delta}} A_0(\Mn^2)  = \frac{-4\pi}{(2\pi R)^{d-2}} 
  \Big[ \int_1^\infty dy \; [{\cal B}^{\delta}(y)-1 ] y^{d/2} \big(1+y^{\delta/2-2}\big) + 
  \frac{2}{d-2} - \frac{2}{d+\delta-2} \Big] \\
{\cal I}_\alpha = \frac{\pi^\alpha}{\Gamma(\alpha)} \big[
 \int_1^\infty dy \; [{\cal B}^{\delta}(y)-1 ] (y^{\alpha-1}+y^{\delta/2-1-\alpha}) 
 -\frac{1}{\alpha} - \frac{1}{\delta/2-\alpha} \big]
\end{gather}
Notice that the sum of the $A_0(\Mn^2)$ function has no $1/\epsilon$ pole. 
The special cases ${\cal I}_0$, ${\cal I}_1$ are needed respectively to evaluate terms $\sum A_0(m^2_{0,1})$
in the results of the previous section and the series of eq.~(\ref{eq:taudiag})
\begin{equation}
\begin{split}
{\cal I}_0 &= -1 \\
{\cal I}_1 &= 
 \pi \big[ \int_1^\infty dy \; [{\cal B}^{\delta}(y)-1 ] (1+y^{\delta/2-2})  - 
 \frac{\delta}{\delta-2} \big]
\end{split}
\end{equation}
Finally, for $(d+\delta)$ {\it even}
\begin{equation}
\begin{split}
F(\bar d,\bar\delta) \frac{\mu^{2\epsilon}}{R^{\bar\delta}} 
 &\sum_{n\in \mathbb{Z}^{\bar\delta}} \Mn^2 B_i (m^2,\Mn^2,m^2) =
  -\frac{\delta F(d,\delta)}{(2\pi)^{d-3}} \int dx \, x^{d/2-2} u_i(x) \\ 
  \Big\{ &\frac{1}{R^{d+\delta-2}} \, f_2(mR,x,d,\delta) + 
  \frac{m^{d+\delta-2}}{\Gamma\left((d+\delta)/2\right)} 
  \left[ -\frac{\pi(1-x)^2}{x} \right]^{(d+\delta)/2-1}  \\
  &\Big( \frac{1}{\epsilon} +\log \frac{\mu^2}{m^2} +\log\frac{2\sqrt{x}}{(1-x)^2} -
  \gamma_E+\frac{1}{F(d,\delta)}\frac{dF}{d\epsilon}\Big|_{\epsilon=0} -\frac{1}{\delta}\Big) \Big\}
\end{split}
\end{equation}
while for $(d+\delta)$ {\it odd}
\begin{equation}
\begin{split}
F(\bar d,\bar\delta) \frac{\mu^{2\epsilon}}{R^{\bar\delta}} 
 &\sum_{n\in \mathbb{Z}^{\bar\delta}} \Mn^2 B_i (m^2,\Mn^2,m^2) =
  -\frac{\delta F(d,\delta)}{(2\pi)^{d-3}} \int dx \, x^{d/2-2} u_i(x) \\ 
  \Big\{ &\frac{1}{R^{d+\delta-2}} \, f_2(mR,x,d,\delta) + 
  m^{d+\delta-2} \, \Gamma\left(1-\frac{d+\delta}{2} \right) 
  \left[ \frac{\pi(1-x)^2}{x} \right]^{(d+\delta)/2-1}  \Big\}
\end{split}
\end{equation}
where $i=0,1$ and $u_0(x)=1$, $u_1(x)=(x-1)$ and we have defined
\begin{equation}
\begin{split}
f_2(mR,x,d,\delta) = 
 &\int_1^\infty dy \; e^{-y\pi a^2(x)} \big[2B^\prime(y) B^{\delta-1}(y)+y^{-(d+\delta)/2} \big]-\\
 &\int_1^\infty dy \; e^{-\pi a^2(x)/y} y^{(d+\delta)/2-2} 
          \big[2yB^\prime(y) B^{\delta-1}(y)+(B(y)-1) B^{\delta-1}(y) \big] 
\end{split}
\end{equation}


\begin{thebibliography}{nn}\small

\bibitem{large} 
N. Arkani-Hamed, S. Dimopoulos and G. Dvali \plb{429}{1998}{263}, \hepph{9803315}; 
\prd{59}{1999}{086004}, \hepph{9807344}.

\bibitem{large2}
I. Antoniadis, N. Arkani-Hamed, S. Dimopoulos and G. Dvali 
\plb{436}{1998}{263}, \hepph{9804398}.

\bibitem{nova} 
S.~Cullen and M.~Perelstein, \prl{83}{1999}{268}, \hepph{9903422}.
%``SN1987A constraints on large compact dimensions,''

\bibitem{GRW} 
G. Giudice, R. Rattazzi and J. D. Wells \npb{544}{1999}{3}, \hepph{9811291}.

\bibitem{KKgr} 
E.A. Mirabelli, M. Perelstein and M.E. Peskin, \prl{82}{1999}{2236}, \hepph{9811337}; 
Z. Kakushadze and S. H. Tye, \npb{548}{1999}{180}, \hepth{9809147}; 
G. Shiu, R. Shrock and S. H. Tye, \plb{458}{1999}{274}, \hepph{9904262}.

\bibitem{KKgr2}
T.~Han, J.~D.~Lykken and R.~Zhang, \prd{59}{1999}{105006}, \hepph{9811350}.

\bibitem{EWPOgr1} 
P.~Das and S.~Raychaudhuri, \hepph{9908205}.

\bibitem{EWPOgr2}
T.~Han, D.~Marfatia and R.~Zhang, \hepph{0001320}.

\bibitem{Bij}
R.~Akhoury and J.~J.~van der Bij, \hepph{0005055}.

\bibitem{grasser} M.L. Graesser, \prd{61}{2000}{74019}, \hepph{9902310}.

\bibitem{STU} M. Peskin and T. Takeuchi, \prl{65}{1990}{194}.

\bibitem{epsilon}
G. Altarelli, R. Barbieri, \plb{253}{1990}{161}. 
For a review see e.g.\ R. Barbieri, CERN-TH.6659/92 (available from the CERN preprint server).

\bibitem{Dono} J.~F.~Donoghue, \prd{50}{1994}{3874}; \grqc{9512024}.
%``Introduction to the Effective Field Theory Description of Gravity,''

\bibitem{sundrum}
R.~Sundrum, \prd{59}{1999}{085009}, \hepph{9805471}; 
A. Dobado and A.L. Maroto, \npb{592}{2001}{203}, \hepph{0007100}.

\bibitem{ibanez}
G.~Aldazabal, S.~Franco, L.~E.~Ibanez, R.~Rabadan and A.~M.~Uranga, \hepph{0011132}.
%``Intersecting brane worlds,''

\bibitem{Aldazabal:2000sa}
G.~Aldazabal, L.~E.~Ibanez, F.~Quevedo and A.~M.~Uranga, \jhep{08}{2000}{002}, \hepth{0005067}.

\bibitem{Antoniadis:2001bh} I.~Antoniadis, \hepth{0102202}.

\bibitem{Rubakov} V.~A.~Rubakov and M.~E.~Shaposhnikov, \plb{125}{1983}{136}.

\bibitem{Dvali} G.~Dvali and M.~Shifman, \plb{396}{1997}{64}, \hepth{9612128}.

\bibitem{pasarin} 
See for instance {\it The Standard Model in the Making}, D. Bardin, G. Passarino,
Clarendon Press, Oxford 1999, at page 196.


\bibitem{charmousis} C. Charmousis, R. Emparan and R. Gregory, \hepth{0101198}.

\bibitem{branoni}
Astrophysical bounds have been studied in
T. Kugo and K. Yoshioka, \npb{594}{2001}{301}, \hepph{9912496}, 
Real and virtual branon effects at colliders have been studied in
P. Creminelli, A. Strumia, \npb{596}{2001}{125}, \hepph{0007267}.

\bibitem{Giudice:2001av}
G.~F.~Giudice, R.~Rattazzi and J.~D.~Wells, \npb{595}{2001}{250}, \hepph{0002178}.

\bibitem{spinors} 
B.S. DeWitt in {\it Relativity, Groups and Topology {\rm I}}, Les Houches 1963, edited by
C. DeWitt and B. DeWitt, Blackie and son limited  1964;
B.S. DeWitt in {\it Relativity, Groups and Topology {\rm II}}, Les Houches 1983, 
edited by B.S. DeWitt and R. Stora, North-Holland 1984.

\bibitem{des} S. Deser, P. van Nieuwenhuizen, \prd{10}{1974}{411}.

\bibitem{woo} R.P. Woodard, \plb{148}{1984}{440}.

\bibitem{tomaras} I.~Antoniadis, J.~Iliopoulos and T.~N.~Tomaras,
%``Gauge Invariance In Quantum Gravity,''
Nucl.\ Phys.\ B {\bf 267}, 497 (1986); D.~A.~Johnston,
%``Sedentary Ghost Poles In Higher Derivative Gravity,''
Nucl.\ Phys.\ B {\bf 297}, 721 (1988);
B.~de Wit and N.~D.~Hari Dass,
%``Gauge independence in quantum gravity,''
Nucl.\ Phys.\ B {\bf 374}, 99 (1992).

\bibitem{Mathematica}
S. Wolfram, {\em The Mathematica book}, 3rd ed.\ (Wolfram media/Cambridge Univ. Press, 1996).

\bibitem{virtual} 
J.L. Hewett, \prl{82}{1999}{4765}, \hepph{9811356}; 
K. Agashe and N.G. Deshpande, \plb{456}{1999}{60}, \hepph{9902263}.

\bibitem{bando} 
M. Bando, T. Kugo, T. Noguchi and K. Yoshioka, \prl{83}{1999}{3601}, \hepph{9906549}, 
J. Hisano, N. Okada, \prd{61}{2000}{106003}, \hepph{9909555}.

\bibitem{nda} S. Weinberg, Physica {\bf 96A}, 327 (1979);
H. Georgi and A. Manhoar, \npb{234}{1984}{189}.

\bibitem{luty} Z.~Chacko, M.~A.~Luty and E.~Ponton, \jhep{07}{2000}{036}, \hepph{9909248}.

\bibitem{NRO} R. Barbieri, A. Strumia, \plb{462}{1999}{144}, \hepph{9905281}.

\bibitem{vincoli}
The OPAL collaboration, \hepex{0005002}; 
comparable bounds have been produced by the
DELPHI, preprint CERN--EP/2000-021,
L3 collaboration, \plb{464}{1999}{135}; 
and ALEPH collaborations.
The D0 collaboration, \prl{86}{2001}{1156}, \hepph{0008065}.
For a review and references, see e.g.\ G. Landsberg, \hepex{0009038}.

\bibitem{g-2} {Muon $g-2$ collaboration}, \hepex{0102017}.
It does not look impossible that the `anomaly' in the $g-2$ is due to
QCD corrections to the photon propagator and/or to
light-by-light scattering.

\bibitem{antico} F.A. Berends and R. Gastmans, \plb{55}{1975}{311}.

\bibitem{Passarino} 
G. 't Hooft, M. Veltman, \npb{153}{1979}{365};
G. Passarino, M. Veltman, \npb{160}{1979}{151}.




\end{thebibliography}
\end{document}